\documentclass[pdflatex,sn-mathphys-num]{sn-jnl}

\usepackage{graphicx}%
\usepackage{multirow}%
\usepackage{amsmath,amssymb,amsfonts}%
\usepackage{amsthm}%
\usepackage{mathrsfs}%
\usepackage[title]{appendix}%
\usepackage{xcolor}%
\usepackage{textcomp}%
\usepackage{manyfoot}%
\usepackage{booktabs}%
\usepackage{algorithm}%
\usepackage{algorithmicx}%
\usepackage{algpseudocode}%
\usepackage{listings}%
\usepackage{physics}

\theoremstyle{thmstyleone}%
\theoremstyle{thmstyletwo}%

\theoremstyle{thmstylethree}%

\makeatletter
\newcommand{\ostar}{\mathbin{\mathpalette\make@circled\star}}
\newcommand{\make@circled}[2]{%
  \ooalign{$\m@th#1\smallbigcirc{#1}$\cr\hidewidth$\m@th#1#2$\hidewidth\cr}%
}
\newcommand{\smallbigcirc}[1]{%
  \vcenter{\hbox{\scalebox{0.77778}{$\m@th#1\bigcirc$}}}%
}
\makeatother

\begin{document}

\title[Article Title]{Let the Quantum Creep In: Designing Quantum Neural Network Models by Gradually Swapping Out Classical Components}


\author*[1]{
\fnm{Peiyong}
\sur{Wang}
}
\email{peiyongw@student.unimelb.edu.au}

\author[3,1]{
\fnm{Casey R.}
\sur{Myers}
}
\email{casey.myers@uwa.edu.au}

\author[2]{
\fnm{Lloyd C. L.}
\sur{Hollenberg}
}
\email{lloydch@unimelb.edu.au}

\author[1]{
\fnm{Udaya}
\sur{Parampalli}
}
\email{udaya@unimelb.edu.au}

\affil[1]{
\orgdiv{School of Computing and Information Systems, Faculty of Engineering and Information Technology},
\orgname{The University of Melbourne},
\orgaddress{
 \country{Australia}}
}

\affil[2]{
\orgdiv{School of Physics},
\orgname{The University of Melbourne},
\orgaddress{
\country{Australia}}
}

\affil[3]{
\orgdiv{School of Physics, Mathematics and Computing},
\orgname{The University of Western Australia},
\orgaddress{
\country{Australia}}
}


\abstract{

Artificial Intelligence (AI), with its multiplier effect and wide applications in multiple areas, could potentially be an important application of quantum computing. Since modern AI systems are often built on neural networks, the design of quantum neural networks becomes a key challenge in integrating quantum computing into AI. To provide a more fine-grained characterisation of the impact of quantum components on the performance of neural networks, we propose a framework where classical neural network layers are gradually replaced by quantum layers that have the same type of input and output while keeping the flow of information between layers unchanged, different from most current research in quantum neural network, which favours an end-to-end quantum model. We start with a simple three-layer classical neural network without any normalisation layers or activation functions, and gradually change the classical layers to the corresponding quantum versions. We conduct numerical experiments on image classification datasets such as the MNIST, FashionMNIST and CIFAR-10 datasets to demonstrate the change of performance brought by the systematic introduction of quantum components. Through this framework, our research sheds new light on the design of future quantum neural network models where it could be more favourable to search for methods and frameworks that harness the advantages from both the classical and quantum worlds.

}

\keywords{Quantum Machine Learning, Quanvolutional Neural Networks, Image Classification}



\maketitle

\section{Introduction}\label{intro}


Machine Learning (ML), or more generally, Artificial Intelligence (AI), aims to develop AI agents and systems that could simulate or even surpass human intelligence. With the promise of quantum computing and quantum advantage in many other fields, such as quantum chemistry and quantum optimisation, there has been rising interest on the application of quantum computing to the research and development of AI and machine learning. However, as of now, with the current state of quantum computing, it is still a contentious issue whether it is truly useful to introduce quantum computing to AI. Some research, originating mainly from the quantum computing community, argues that quantum computing has advantages in machine learning and AI, such as speed-up for both statistical machine learning algorithms \cite{Biamonte2017-pg} and modern deep neural networks \cite{Guo2024-ph}, as well as memory advantages in sequence learning tasks \cite{Anschuetz2023-mk, Anschuetz2024-lv}. There are also some doubts about the current research paradigm of combining quantum computing and AI to harness quantum advantage for both runtime acceleration and/or performance boost, ranging from whether the conventional notion of quantum advantage is the right goal for quantum machine learning \cite{Schuld2022-ll}, to results showing that quantum machine learning models rarely outperform the corresponding off-the-shelf classical machine learning models \cite{Bowles2024-kb}. Most of the time, the quantum machine learning models studied in \cite{Bowles2024-kb}, such as quantum neural networks (QNN), perform poorly compared to the classical multilayer perceptron and simple convolutional neural network.

Unlike quantum kernel machines \cite{Havlicek2019-wn}, which replace the classical kernel function with kernel functions calculated via the evaluation of quantum circuits, the correspondence between classical and quantum neural networks is less straightforward. Classical neural networks (NN) have a clear hierarchical layered structure. Different layers often deal with different levels of features. For example, in a convolutional neural network for image classification, layers close to input data learn low-level features such as shapes and edges, while layers close to the output layer learn high-level features related to semantic concepts in the images \cite{zeiler2014visualizing}. However, most quantum neural networks in today's research lack such hierarchical structure, especially for those that follow a sandwich ``Data Encoding $\rightarrow$ Quantum Process $\rightarrow$ Results Readout" structure. From the perspective of classical neural network architectures, no matter how many ``quantum layers" are there in the middle of the sandwich, it is still a single layer since all those quantum layers could be represented with a single linear map. This leads to an inherent disadvantage when directly comparing most of the quantum neural network models in current research with the corresponding off-the-shelf classical neural networks.

In this paper, we explore these issues by focussing on the transition from a fully classical model to a classical-quantum hybrid model, \verb+HybridNet+, in which the layers are realised via (simulated) quantum circuits, but the information flow between layers remains classical. As our main contribution, we propose such a gradual transition strategy for benchmarking the effectiveness of quantum neural network layers for particular tasks. We proposed a novel trainable quanvolution \cite{Henderson2020-yw} kernel, \verb+FlippedQuanv3x3+, based on the flipped model for quantum machine learning \cite{Jerbi2023-rg}. We also adopt the data reuploading circuit \cite{Perez-Salinas2020-gi}, combined with the Hamiltonian embedding of classical data \cite{QuantumDataEncodingFromBook-Schuld2021-ak, Yang2023-pc}, as introduced in \cite{wang2024quantumhamiltonianembeddingimages}, \verb+DataReUploadingLinear+, to mimic the effect of a classical linear (dense) layer.

This paper is organised as follows. In Section~\ref{sec:methods}, we introduce the framework of the gradual transition strategy with a simple classical neural network as an example. We also introduce the details for the implementations of \verb+FlippedQuanv3x3+ and \verb+DataReUploadingLinear+ in Section~\ref{sec:methods}. In Section~\ref{sec:experiments}, we investigate the performance of our hybrid model by numerical experiments on three famous image classification datasets: MNIST, FashionMNIST and CIFAR-10, and analyse the results. We discuss the general implications of our results and possible future directions in Section~\ref{sec:discussion}.

\section{\label{sec:methods}Methods}

\subsection{Let the Quantum Creep In}
\begin{figure}[]
    \centering
    \includegraphics[width=0.85\textwidth]{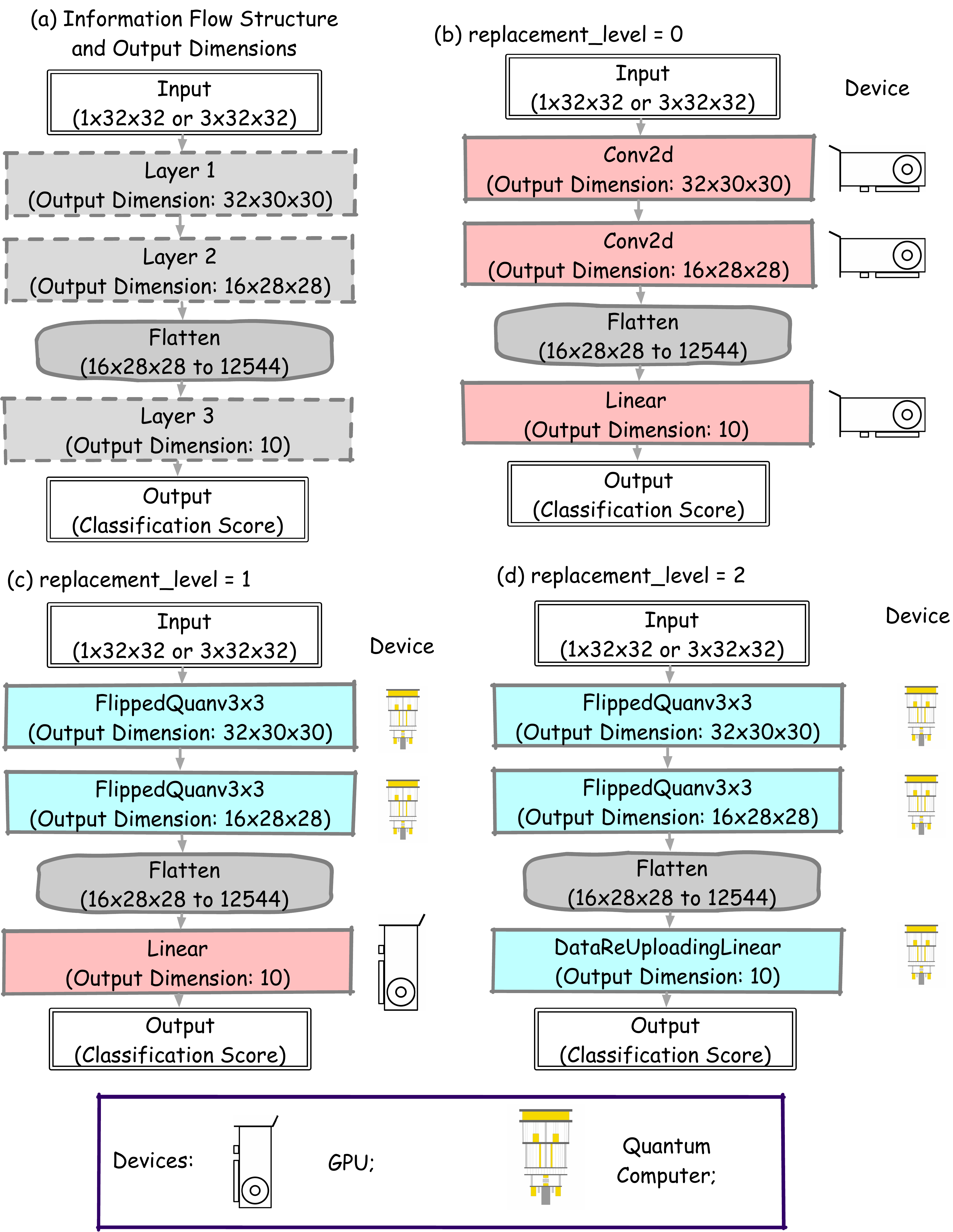}
    \caption{Overview of the framework proposed in this paper. The symbol for the quantum computer is inspired by \cite{cerezo2024does}. (a) the information flow structure and the required dimensions of the input and output of each vacancy for candidate neural network layers. Double-lined boxes are the input and output of the neural network; Dash-lined boxes are layer vacancies for candidate neural network layers. Alongside the block of layers are the devices where the layer operation will mainly be executed on. The information passed between layers and the flatten operation are classical, while the candidate neural network layers could be either classical or (simulated) quantum. (b) The hybrid neural network, \texttt{HybridNet}, when \texttt{replacement\_level = 0}. In this case all vacancies are filled with classical neural network layers (\texttt{Conv2d} and \texttt{Linear}). All these layers are executed on a GPU with classical neural network libraries. (c) \texttt{HybridNet}, when \texttt{replacement\_level = 1}. In this case, the classical convolution layers \texttt{Conv2d} are replaced with its quantum counterpart, \texttt{FlippedQuanv3x3}, while the classical \texttt{Linear} layer left unchanged. The two quantum layers could be executed either via GPU simulation or on an actual quantum device. In this paper, they are simulated on a GPU since the current accessibility of quantum processors prohibits us from executing a very large number of circuits. (d) \texttt{HybridNet}, when \texttt{replacement\_level = 2}. In this case, all the classical layers in (b) are replaced with their quantum counterpart, i.e.  \texttt{Conv2d}$\rightarrow$ \texttt{FlippedQuanv3x3} and \texttt{Linear}$\rightarrow$\texttt{DataReUploadingLinear}. All quantum layers are simulated on a GPU when training and testing the neural network model.}
    \label{fig:fig1}
\end{figure}

\begin{figure}[ht!]
    \centering
    \includegraphics[width=0.85\textwidth]{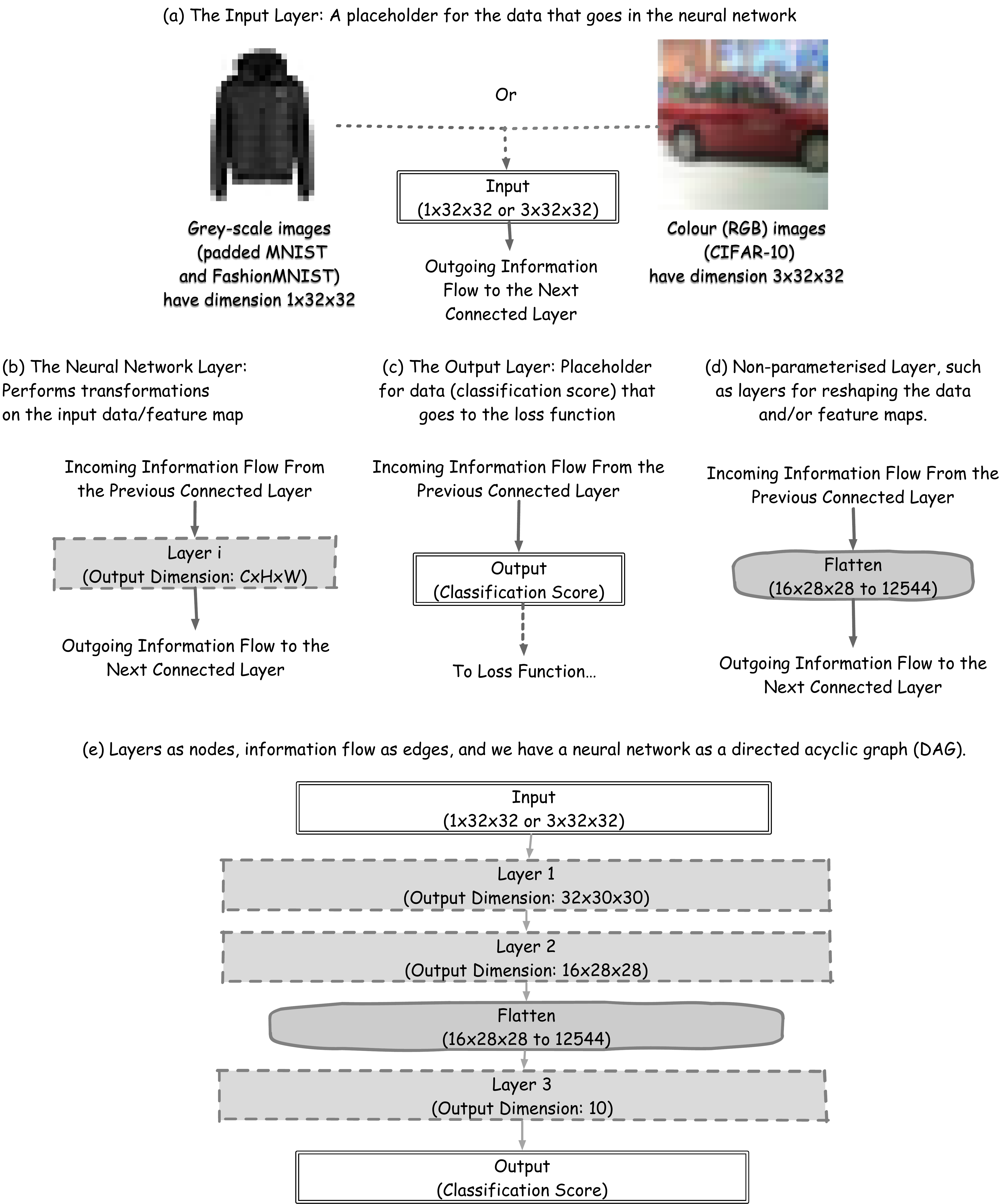}
    \caption{A more detailed account of the components in the neural network architecture shown in Fig.~\ref{fig:fig1}. Neural networks are essentially directed acyclic graphs, with layers as nodes and the flow of information as directed edges. The input layer (a) is just a placeholder for the input data. Grey-scale images from MNIST and FashionMNIST have only one channel, so the dimension is $1\times 32\times 32$ (after padding with zero); Colour images from CIFAR-10 have three channels, so the dimension is $3\times 32\times 32$. For the trainable layers (b), each has a required input dimension and an output dimension determined by the hyper-parameters of the layer. The dimension (shape) of the incoming data is the same as the output of the previous layer. The dimension of the outgoing data (feature map) is $C(\text{hannel})\times H(\text{eight})\times W\textrm{(idth)}$ for \texttt{Conv2d} and \texttt{FlippedQuanv3x3} layers. For \texttt{Linear} and \texttt{DataReUploadingLinear} layers, it is a number of the vector dimension. Both \texttt{Conv2d}-like and \texttt{Linear}-like layers have hyper-parameters that could control the behaviour of the layer and change the dimension of the output. 
    The output layer (c) is also a placeholder for the information that is going to the loss function. A neural network also contains non-trainable layers such as the flatten layer (d), which reshapes the multi-channel feature map from a convolution/quanvolution layer to a 1D vector. Putting all these together with layers as nodes and information flow as directed edges, we have the architecture for a neural network (e).}
    \label{fig:fig2}
\end{figure}

Most of the quantum machine learning models, especially quantum neural networks, are end-to-end quantum. These QNN architectures can be written in the form of a parameterised unitary
\begin{equation}\label{eqn:conventional-model}
    f(x;\boldsymbol{\theta}) = \bra{x}\mathcal{O}(\boldsymbol{\theta})\ket{x},
\end{equation}
where $\ket{x}$ is the quantum encoding of classical input data $x$, $\mathcal{O}(\boldsymbol{\theta})$ is an observable parameterised by $\boldsymbol{\theta}$, formulated by a unitary quantum circuit parameterised by $\boldsymbol{\theta}$ absorbed into some cost-function related observable $\mathcal{O}$. This can also be formulated in the form of a quantum channel (such as the quantum convolutional neural network \cite{QCNN-Cong_2019}):
\begin{equation}\label{eqn:qcnn-model}
    f(x;\boldsymbol{\theta}) = \operatorname{Tr}[\mathcal{O}\rho(x;\boldsymbol{\theta})],
\end{equation}
where
\begin{equation}
    \rho(x;\boldsymbol{\theta}) = \sum_i K_i({\boldsymbol{\theta}})\rho_{{x}} K_i({\boldsymbol{\theta}})^\dagger,
\end{equation}
$\rho_{{x}}$ is the quantum encoding of classical input data $x$ and $\sum_i K_i({\boldsymbol{\theta}})^\dagger K_i({\boldsymbol{\theta}}) = I$ are the Kraus operator representation of the quantum channel parameterised by ${\boldsymbol{\theta}}$. 

Both of these QNN architectures lack the hierarchical layered structure that commonly exists in classical neural networks, giving these models a major disadvantage compared to off-the-shelf classical neural network models. Hence, it would be hard to determine whether the lower performance of QNN models compared to classical NN models is due to the absence of a layered structure, or the intrinsic ineffectiveness of the quantum model. 

To address this problem, we adopt an approach in which we gradually replace the layers of classical neural networks with quantum layers that mimic the behaviour of their classical counterparts (see Fig.~\ref{fig:fig1}). Fig.~\ref{fig:fig2} offers a detailed legend of the information-passing structure adopted in this paper, for both classical and quantum layers. We dub this approach ``let the quantum creep in". A general requirement for the quantum layers that are replacing the original classical layers is that they need to share the same input / output data type as the classical layers. For example, the classical (two-dimensional) convolutional layer takes a (set of) 2-D feature map(s) to another (set of) 2-D feature map(s). It usually has the following specifications: number of input channels, number of output channels, size of the kernel, stride (step size) of the convolution operation, and padding specifications of the input image / feature map. The quantum replacement should also have the same input / output data types and the same specifications, although in the implementations we fix the specifications besides the numbers of input / output channels. This is also the case for the quantum replacement of the linear layer: it should also take a vector, usually a flattened feature map from the previous convolution layer, and produce another vector. For convenience in the implementation, we fix the dimensions of the input and output vectors of the quantum ``linear" layer. Keeping the input / output data types unchanged gives us minimal disturbance to the flow of information in the neural network. In this case, the performance change can be more confidently attributed to the change in neural network layers.

Since some choices of classical activation functions and layer normalisation could introduce a bias that makes the neural network more suitable for real-world data \cite{Teney2024-wy}, the ReLU (rectified linear unit) activation function, as well as batch / layer normalisation, are removed from the fully classical model (\texttt{replacement\_level = 0}), which is the baseline for our study. 

\subsection{\label{sec:flipped-quanv}Flipped Quanvolution}

The flipped model was first proposed in \cite{Jerbi2023-rg}:
\begin{equation}\label{eqn:flipped-model}
    f_{\boldsymbol{\theta}} (x) = \operatorname{Tr}[\rho(\boldsymbol{\theta})\mathcal{O}(x)],
\end{equation}
where $\boldsymbol{\theta}$ is the set of trainable parameters and $x$ is the input data. It exchanges the position of the input data $x$ and trainable parameters $\boldsymbol{\theta}$ compared to the common form in Eqn.~\ref{eqn:conventional-model}, shifting the data to the observable side and the parameters to the initial state side. If the parameterised initial state is a pure state, i.e. $\rho(\boldsymbol{\theta}) = \ket{\varphi(\boldsymbol{\theta})}\bra{\varphi(\boldsymbol{\theta})}$, then $f_{\boldsymbol{\theta}}(x)$ becomes
\begin{equation}
    f_{\boldsymbol{\theta}}(x) = \bra{\varphi(\boldsymbol{\theta})}\mathcal{O}(x)\ket{\varphi(\boldsymbol{\theta})}.
\end{equation}
For the two-qubit case, if we let
\begin{equation}
    \ket{\varphi(\boldsymbol{\theta})} = \begin{bmatrix}
a \\
b \\
c \\
d \\
\end{bmatrix},
\end{equation}
where $a, b, c, d \in \mathbb{C}$, and
\begin{equation}\label{eqn:patch-to-hermitian}
    \mathcal{O}(x) = \frac{1}{2}(M_x+M^T_x),
\end{equation}
where $M$ is the padded image patch and $\mathcal{O}$ is now guaranteed to be Hermitian. We fix the kernel size for the \verb+FlippedQuanv3x3+ layer to three by three, which means that there will be nine different pixel values (or feature map values) $x_1, x_2, \cdots, x_8, x_9$ in the kernel view. Then we have
\begin{equation}
    M_x = \begin{bmatrix}
        x_1 & x_2 & x_3 & 0 \\
        x_4 & x_5 & x_6 & 0 \\
        x_7 & x_8 & x_9 & 0 \\
        0 & 0 & 0 & 0
    \end{bmatrix}.
\end{equation}
So
\begin{equation}
     \mathcal{O}(x) = \frac{1}{2}\begin{bmatrix}
         2x_1 & x_2+x_4 & x_3+x_7 & 0 \\
        x_2+x_4 & 2x_5 & x_6+x_8 & 0 \\
        x_3+x_7 & x_6+x_8 & 2x_9 & 0 \\
        0 & 0 & 0 & 0
     \end{bmatrix}.
\end{equation}
Then we have the quantum function for the flipped quanvolution with single-channel input:
\begin{equation}
    \begin{split}
        f_{\boldsymbol{\theta}}(x) &= \bra{\varphi(\boldsymbol{\theta})}\mathcal{O}(x)\ket{\varphi(\boldsymbol{\theta})}\\
        &=\frac{1}{2}\begin{bmatrix}
            a^\star & b^\star & c^\star & d^\star
        \end{bmatrix}
        \begin{bmatrix}
         2x_1 & x_2+x_4 & x_3+x_7 & 0 \\
        x_2+x_4 & 2x_5 & x_6+x_8 & 0 \\
        x_3+x_7 & x_6+x_8 & 2x_9 & 0 \\
        0 & 0 & 0 & 0
     \end{bmatrix}
     \begin{bmatrix}
a \\
b \\
c \\
d \\
\end{bmatrix}\\
&=\frac{1}{2} [a^\star (2 a x_1+b x_2+b x_4+c x_3+c x_7)+b^\star (a x_2+a x_4+2 b x_5+c x_6+c x_8)\\ 
&\quad+ c^\star (a x_3+a x_7+b x_6+b x_8+2 c x_9)]\\
&= a a^\star x_1 + \frac{1}{2} (b a^\star+a b^\star)x_2 + \frac{1}{2}(c a^\star+a c^\star)x_3 + \frac{1}{2} (b a^\star+a b^\star) x_4\\
&\quad+ b  b^\star x_5 + \frac{1}{2} (c b^\star+b c^\star) x_6+ \frac{1}{2} (c a^\star+a c^\star) x_7 + \frac{1}{2}  (c b^\star+b c^\star)x_8 + c  c^\star x_9.
\end{split}
\end{equation}
Compared to the classical convolution function (in a convolutional neural network)
\begin{equation}
    w\ostar x = \sum_{i=1}^9 w_i x_i
\end{equation}
between a three-by-three kernel (weight) matrix
\begin{equation}
    w = \begin{bmatrix}
        w_1 & w_2 & w_3\\
        w_4 & w_5 & w_6\\
        w_7 & w_8 & w_9
    \end{bmatrix},
\end{equation}
and the image patch within the kernel view
\begin{equation}
    x = \begin{bmatrix}
         x_1 & x_2 & x_3  \\
        x_4 & x_5 & x_6  \\
        x_7 & x_8 & x_9
    \end{bmatrix},
\end{equation}
we can see that we have a one-to-one correspondence:
\begin{equation}
    \begin{split}
        w_1 &= a a^\star \\
        w_2 &= \frac{1}{2} (b a^\star+a b^\star)\\
        w_3 &= \frac{1}{2}(c a^\star+a c^\star)\\
        w_4 &= \frac{1}{2} (b a^\star+a b^\star)\\
        w_5 &= b  b^\star\\
        w_6 &= \frac{1}{2} (c b^\star+b c^\star) \\
        w_7 &= \frac{1}{2} (c a^\star+a c^\star) \\
        w_8 &= \frac{1}{2}  (c b^\star+b c^\star) \\
        w_9 &= c  c^\star
    \end{split}
\end{equation}
This one-to-one correspondence ensures that our \texttt{FlippedQuanv3x3} will have a similar effect to the classical convolution layer \texttt{Conv2d}. Unlike the original quanvolutional neural network proposed in \cite{Henderson2020-yw}, in which the quanvolutional layer acts as a random feature extractor for the classical convolutional and linear layers, our proposed method has trainable parameters. Compared to the NNQE model proposed in \cite{Riaz2023-iy}, our data encoding method does not impose a trigonometric bias on the input data, following the design principles introduced in \cite{wang2024quantumhamiltonianembeddingimages}.

\subsubsection{Circuit Implementation}

To prepare the parameterised two-qubit quantum state $\ket{\varphi(\boldsymbol{\theta})}$, we adopt the SU(N) unitary for convenience in implementation:

\begin{equation}
    \mathrm{SU}(\text{N})(\boldsymbol{\theta}) = \exp (\sum_{i=1}^m \mathrm{i} \theta_i G_i), 
\end{equation}
where $m=4^n-1, \; N = 2^n$, $n$ is the number of qubits in the circuit, and $G_i\in\{I, X, Y, Z\}^{\otimes n}\backslash \{I^{\otimes n}\}$, $\boldsymbol{\theta} = \{\theta_1, \cdots, \theta_{4^n-1}\}$. For $n=2$ ($N=4$), $\boldsymbol{\theta}$ is a 15-dimensional vector. $\ket{\varphi(\boldsymbol{\theta})}$ is obtained via applying the SU(4) unitary to the $\ket{00}$ state:

\begin{equation}
    \ket{\varphi(\boldsymbol{\theta})} = \operatorname{SU(4)}(\boldsymbol{\theta})\ket{00}.
\end{equation}

\subsubsection{Multi-Channel Inputs and Outputs}

\begin{figure}[ht!]
    \centering
    \includegraphics[width=0.85\textwidth]{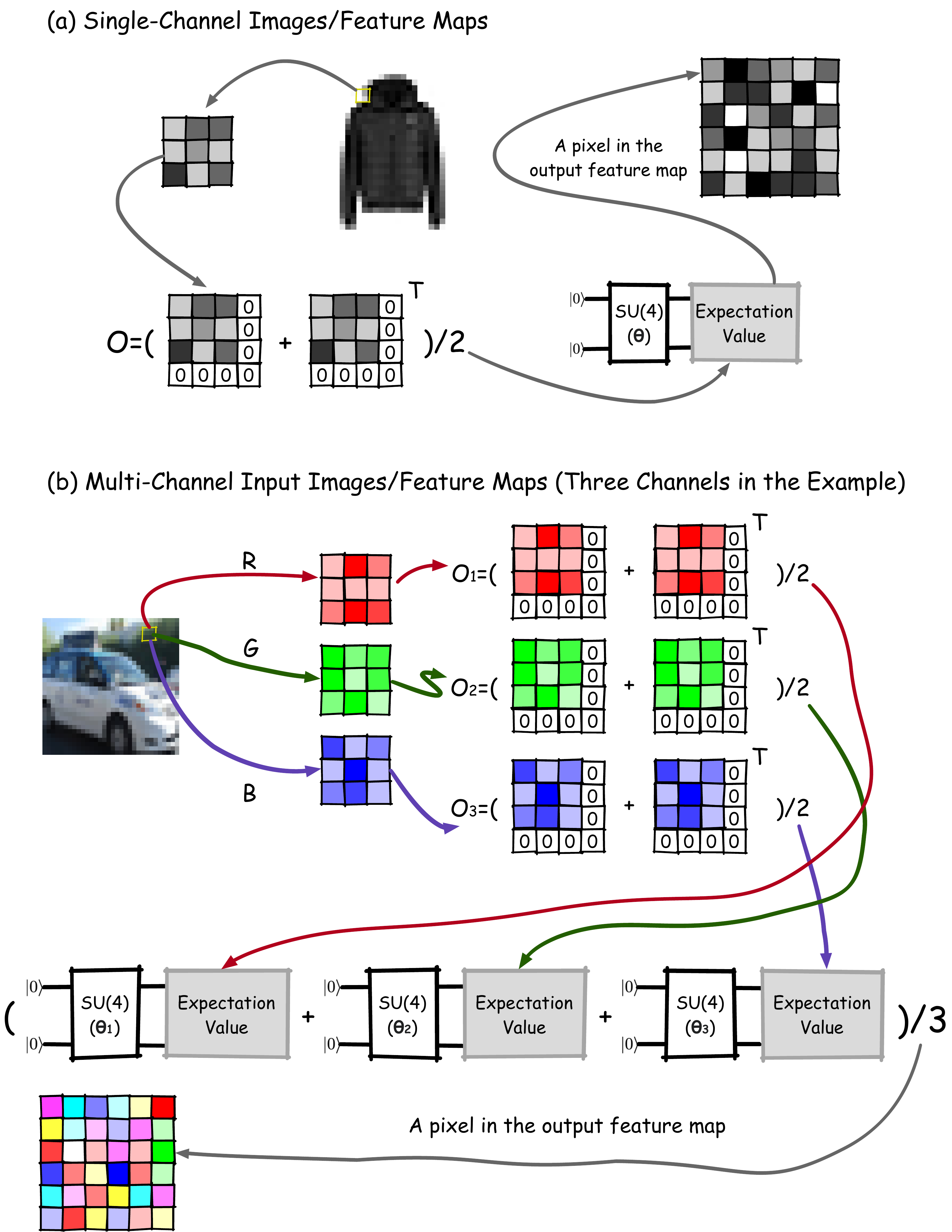}
    \vspace*{3mm}
    \caption{(a) Images and feature maps that only have one channel only need a single circuit for each patch $x$. (b) For images and feature maps that have multiple channels, the patch of image within the view of the \texttt{FlippedQuanv3x3} kernel is a 3-D tensor with the the shape $(C, 3, 3)$. In this example, we take $C=3$, and in this case, three circuits with different parameters are required from the observables constructed from each channel to calculate the output of the \texttt{FlippedQuanv3x3} kernel operation.}
    \label{fig:single-multi-channel-quanv-circs}
\end{figure}

Multi-channel images and feature maps have a shape $(C, H, W)$, where $C$ is the number of channels, $H$ and $W$ are the height and width of the input, respectively. The 3-D patch $X$ within the view of a three-by-three kernel would have the shape $(C, 3, 3)$. We slice the $(C, 3, 3)-$shaped tensor along the channel axis corresponds to a list of patches of shape $(3,3)$: $X = [x^1, x^2, \cdots, x^C]$. We then apply the flipped quanvolution function to each of these slices. However, for each slice, we will have a different set of parameters $\boldsymbol{\theta}^i$. Then, the quanvolution is calculated as
\begin{equation}
    F_{\boldsymbol{\Theta} = [\boldsymbol{\theta}^1, \cdots, \boldsymbol{\theta}^C]}(X) = \frac{1}{C}\sum_{i=1}^C f_{\boldsymbol{\theta}_i}(x^i).
\end{equation}
If the \verb+FlippedQuanv3x3+ needs to output a multichannel feature map, we can just repeat the flipped quanvolution operation with different $\boldsymbol{\Theta}$ on the same input and stack the output feature maps. The single- and multi-channel versions of the \verb+FlippedQuanv3x3+ operation are also depicted in Fig.~\ref{fig:single-multi-channel-quanv-circs}. Note that in the simulation, we also have a bias term added to the output of the \verb+FlippedQuanv3x3+ operation, same as the classical \verb+Conv2d+ layer.

\subsection{\label{sec:data-reuploading-linear}Data Reuploading with Quantum Hamiltonian Embedding}

\begin{figure}[ht!]
    \centering
    \includegraphics[width=0.8\textwidth]{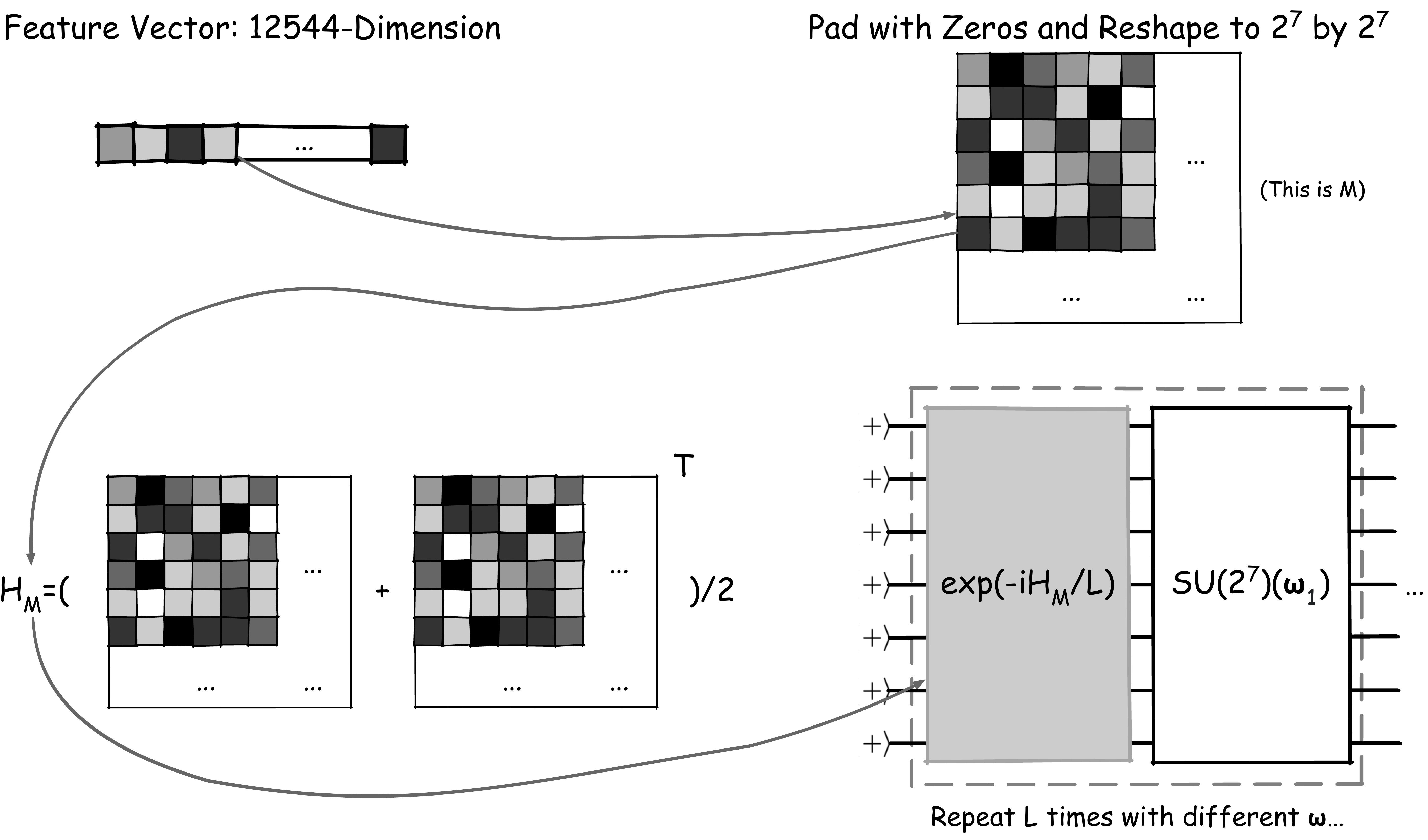}
    \caption{The \texttt{DataReUploadingLinear} layer at the end of the hybrid neural network. It takes a 12544-dimension feature vector from the \texttt{Flatten} layer, pad it with zeros and reshape it to a $2^7\times 2^7$ square matrix $M$. A quantum Hamiltonian $H_M$ is constructed with $M$, and this Hamiltonian will be used to construct the time-evolution operator $W = e^{-i H_M/L}$, where $L$ is the number of layers of the data re-uploading circuit.}
    \label{fig:datareuploadinglinear}
\end{figure}

The replacement for the classical linear (dense) layer, \verb+DataReUploadingLinear+, is designed following the method proposed in \cite{wang2024quantumhamiltonianembeddingimages}, which involves a data reuploading circuit with quantum Hamiltonian embeddings, as shown in Fig~\ref{fig:datareuploadinglinear}. Specifically, since the shape of the output feature map before the linear layer, classical or quantum, is fixed to $16\times 28\times 28$, as shown in Fig.~\ref{fig:fig1}, the flattened feature map, which has dimension 12544, can be padded to $16384 = 4^7 = (2^7)^2$ with zeros. Then the padded feature map is reshaped into a matrix $M$ with dimension $2^7\times 2^7$. As in Eqn.~\ref{eqn:patch-to-hermitian} a Hermitian matrix, $H_M$, could be constructed with the real-valued square matrix $M$ by

\begin{equation}
    H_M = \frac{M+M^T}{2}.
\end{equation}

Then the data embedding unitary can be written as

\begin{equation}
    W(M;t) = e^{\frac{-{i}H_M t}{2}}.
\end{equation}

However, unlike in \cite{wang2024quantumhamiltonianembeddingimages}, where $t$ is a trainable parameter, we fix $t$ to $\frac{1}{L}$ for convenience, where $L$ is the number of layers (or repetitions) of the data reuploading circuit. In the simulation, we fix $L$ to 10.

The parameterised segments of the \verb+DataReUploadingLinear+ are SU($2^7$) unitaries, parameterised by different parameters $\boldsymbol{\omega}_i,\; i\in \{1,2,3,\cdots, L\}$. Then, the state before the measurement can be written as

\begin{equation}
    \ket{\psi(M;\boldsymbol{\Omega})}=\left\{\prod_{i=1}^L[\operatorname{SU}(2^7)(\boldsymbol{\omega}_i)W(M;\frac{1}{L})]\right\}\ket{+}^{\otimes 7},
\end{equation}
where $\boldsymbol{\Omega} = \{\boldsymbol{\omega}_1, \boldsymbol{\omega}_2, \cdots, \boldsymbol{\omega}_L \}$ and $L=10$.

The purpose of the linear layer, both the quantum \verb+DataReUploadingLinear+ and the classical one, is to project the flattened feature map from the previous convolution (or quanvolution) layer to a 10-dimensional space for classification. In \verb+DataReUploadingLinear+, we measure the ten projection operators, $P_i = \ket{i}\bra{i}, i\in \{0, 1, 2, 3, \cdots, 9\}$. Since the maximum value of $i$ is 9, $P_i$ could be constructed as a 4-qubit operator, $\ket{0000}\bra{0000}, \ket{0001}\bra{0001}, \cdots, \ket{1000}\bra{1000}, \ket{1001}\bra{1001}$. We denote the output vector of the last layer as $\boldsymbol{p}$:
\begin{equation}
    \boldsymbol{p} = \begin{bmatrix}
        p_0\\
        p_1\\
       \cdot\\
       \cdot \\
       \cdot \\
        p_8\\
        p_9
    \end{bmatrix}.
\end{equation}

Then for each element $p_i$ in $\boldsymbol{p}$, we have
\begin{equation}
    p_i = \bra{\psi(M;\boldsymbol{\Omega})}(P_i\otimes \boldsymbol{I}_{2^3})\ket{\psi(M;\boldsymbol{\Omega})}.
\end{equation}

$\boldsymbol{p}$ is the ``classification score" of the input image. The index (0, 1, $\cdots$, 8, 9) of the largest element in $\boldsymbol{p}$ is chosen as the predicted label.

Note that, similar to the implementation of the \verb+FlippedQuanv3x3+ layer, we also have a classical bias term added to the measurement results.

\subsection{Loss Function}
We adopt the commonly used cross-entropy loss as our minimisation target during training. Denote $\hat{y}_i$ as the true probability of the input image for label $i$, which is 1 for the true label and 0 for the rest; and $y_{pred}^i$ as the classification score of label $i$ (in our case it equals to $p_i$), then the Cross Entropy function can be written as
\begin{equation}
    \operatorname{CrossEntropy} = -\sum_{i=0}^9 \hat{y}_i \log_2 y_{pred}^i
\end{equation}
for 10-class classification. Since the value of $y_{pred}^i$ could be 0 or a negative number (in the classical case), it is common practice to use $\operatorname{Softmax}(y_{pred}^i)$ to replace $y_{pred}^i$ in the cross-entropy loss \cite{prince2023understanding}:
\begin{equation}
    \operatorname{Softmax}(y_{pred}^i) = \frac{\exp (y_{pred}^i)}{\sum_k \exp(y_{pred}^k)}.
\end{equation}

\section{\label{sec:experiments}Experiments and Results}

In this paper, we trained all three different replacement levels on three different datasets. Each training combination (replacement level $\times$ dataset) is repeated five times with different parameter initialisations. The hyper-parameters are the same throughout all training combinations, and listed in Table~\ref{tab:hyperparam}. We used linear algebra functionalities in PyTorch \cite{pytorch2} for circuit simulation, as well as the Adam optimiser \cite{Adam} implementation in PyTorch, \verb+torch.optim.Adam+, for training. The numerical experiments are conducted on a single NVIDIA H100 GPU.

\begin{table}[ht]
\caption{\label{tab:hyperparam} Training hyperparameters. The rest of the hyperparameters in the classical optimiser are kept the same as the default setting.}
\begin{tabular}{cccc}
\toprule
\textbf{Batch Size} & \textbf{Initial Learning Rate} & \textbf{Number of Iterations} &  \textbf{Optimiser} \\
\midrule
200 & $5\times 10^{-4}$ & 50 & \verb+torch.optim.Adam+\\
\botrule
\end{tabular}
\end{table}

\subsection{Datasets}
\begin{figure}[ht!]
    \centering
    \includegraphics[width=\textwidth]{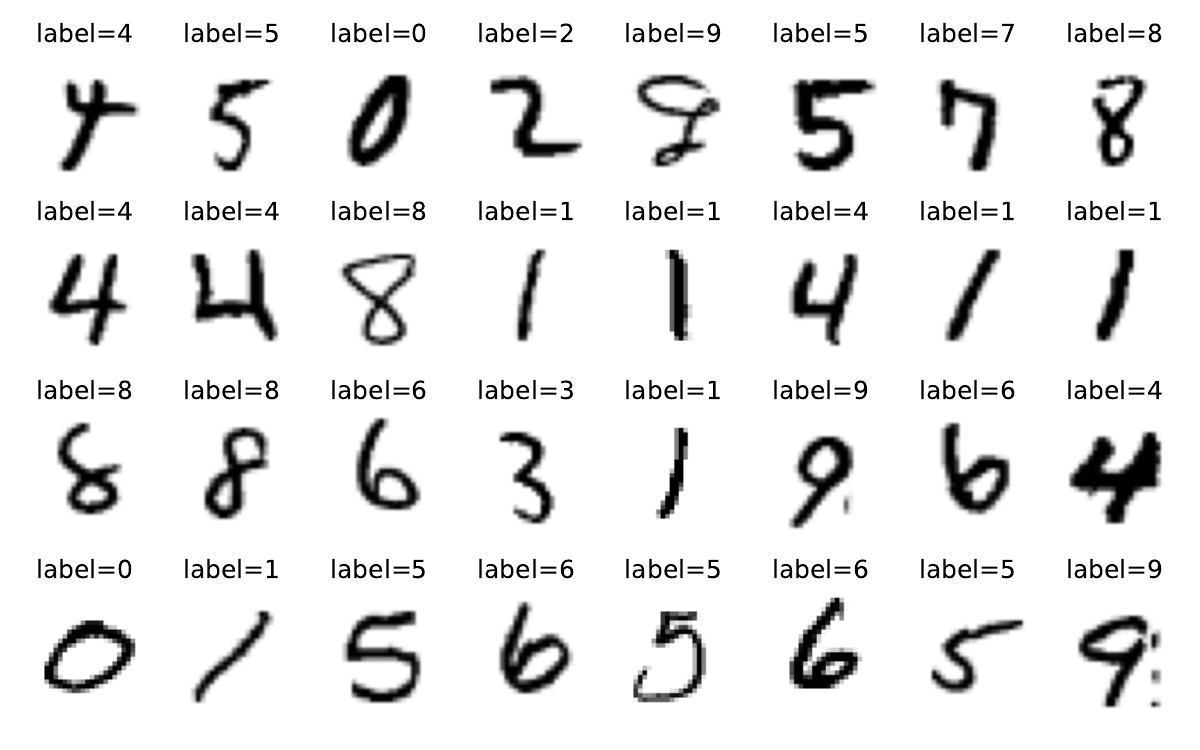}
    \caption{Sample images from the MNIST dataset. The size of the original images is 28 by 28. Images are padded with zeros to 32 by 32.}
    \label{fig:mnist}
\end{figure}

We trained and tested our models on three different datasets: MNIST \cite{lecun2010mnist}, FashionMNIST \cite{xiao2017fashionmnist} and CIFAR-10 \cite{Krizhevsky09learningmultiple}. All three data sets are obtained via Torchvision \cite{torchvision2016}. Both MNIST and FashionMNIST consist of 70000 $28\times 28$ black-and-white images, split into a train set with 60000 images and a test set of 10000 images. The MNIST dataset consists of handwritten digits (0 to 9, see Fig.~\ref{fig:mnist}) extracted from two NIST databases. 

\begin{figure}[ht!]
    \centering
    \includegraphics[width=\textwidth]{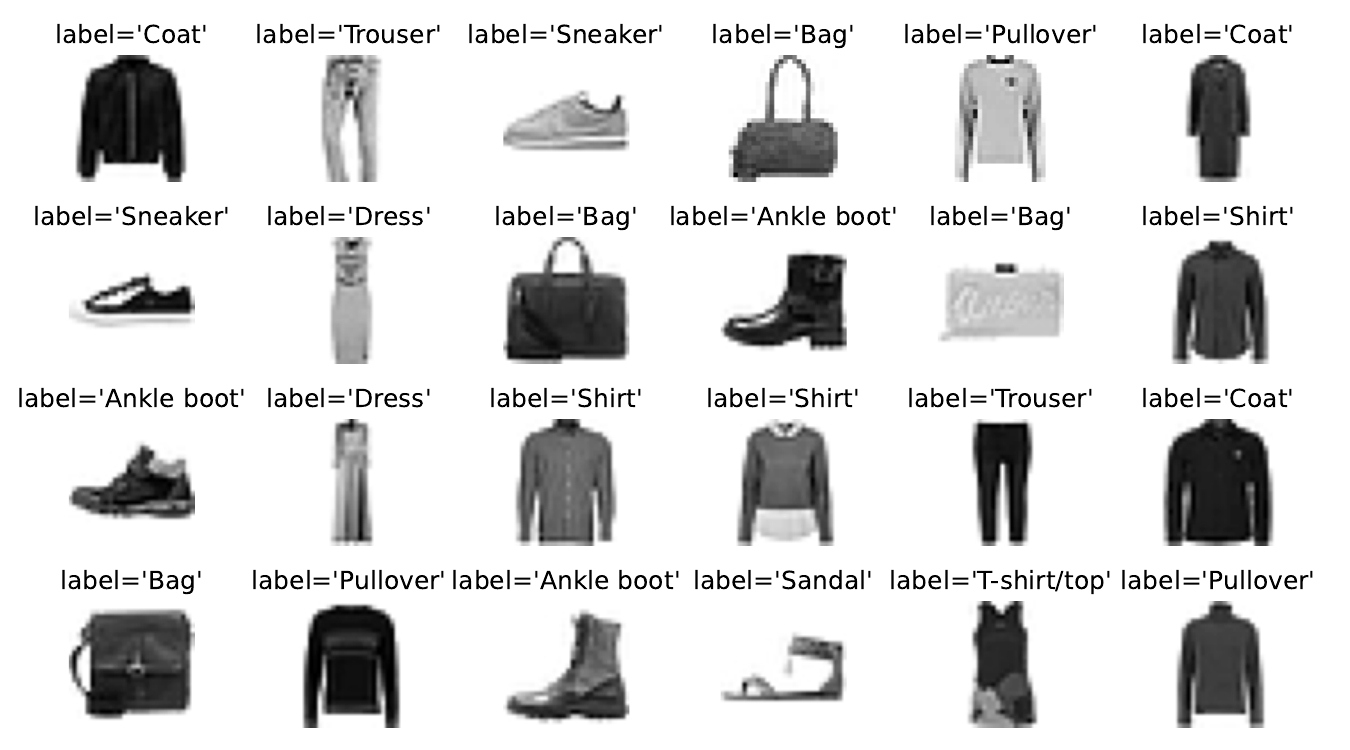}
    \caption{Sample images from the FashionMNIST dataset. The size of the original images is 28 by 28. Images are padded with zeros to 32 by 32.}
    \label{fig:fashionmnist}
\end{figure}

The FashionMNIST is a dataset of Zalando's article images (see Fig.~\ref{fig:fashionmnist}). It was intended as a direct drop-in replacement of the MNIST dataset. Since the classes of images in the FashionMNIST dataset are not integers, they need to be mapped to integer indices of elements in the output $\boldsymbol{p}$ of the neural network (Table~\ref{tab:fashionmnist-label}).

\begin{table}[ht]
\caption{\label{tab:fashionmnist-label} Map between class names and classification indices ($i$ in $p_i$) for the FashionMNIST dataset.}
\tabcolsep=2.1pt
\begin{tabular}{lcccccccccc}
\toprule
\textbf{Class Name} & T-shirt/top & Trouser & Pullover & Dress & Coat & Sandal & Shirt & Sneaker & Bag & Ankle boot \\
\midrule
\textbf{Index}  & 0 & 1 & 2 & 3 & 4 & 5 & 6 & 7 & 8 & 9  \\
\botrule
\end{tabular}
\end{table}

The CIFAR-10 dataset consists of 60000 $32\times 32$ colour images, split into a training set of 50000 images and a test set of 10000 images. There are 10 different classes in the dataset, each mapped to an integer index for the elements $p_i$ in the neural network output  $\boldsymbol{p}$ (see Table~\ref{tab:cifar10-label}).

\begin{figure}[ht!]
    \centering
    \includegraphics[width=\textwidth]{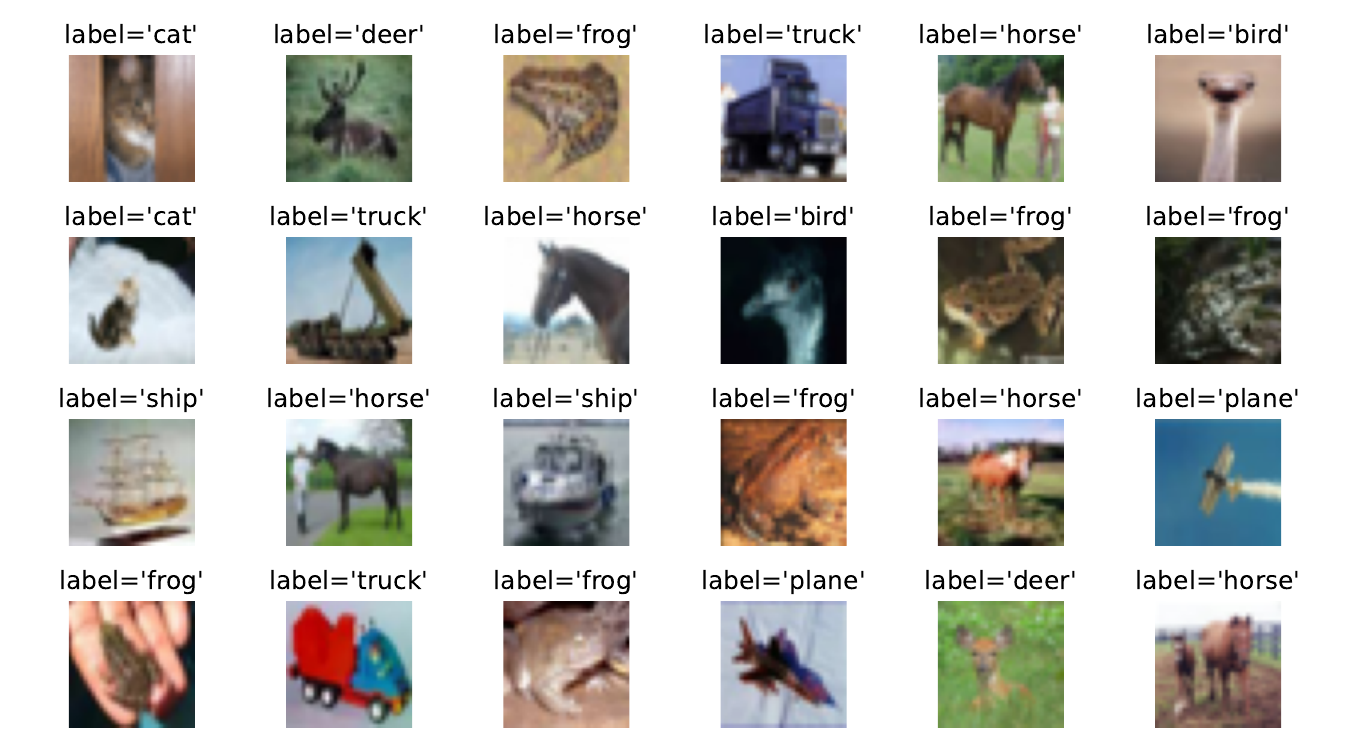}
    \caption{Sample images from the CIFAR-10 dataset. Unlike the images in MNIST and FashionMNIST, images in the CIFAR-10 dataset have three channels: red, green and blue.}
    \label{fig:cifar10}
\end{figure}

\begin{table}[ht]
\caption{\label{tab:cifar10-label} Map between class names and classification indices ($i$ in $p_i$) for the CIFAR-10 dataset.}
\tabcolsep=2.1pt
\begin{tabular}{lcccccccccc}
\toprule
\textbf{Class Name} & plane & car & bird & cat & deer & dog & frog & horse & ship & truck \\
\midrule
\textbf{Index}  & 0 & 1 & 2 & 3 & 4 & 5 & 6 & 7 & 8 & 9  \\
\botrule
\end{tabular}
\end{table}

All images in all three datasets are normalised (for MNIST and FashionMNIST, after the padding from 28 by 28 to 32 by 32) via \texttt{torchvision.transform.Normalize}. \texttt{torchvision.transform.Normalize} takes two major arguments: \texttt{mean} and \texttt{std}. The normalisation process is 
\begin{equation}
    \texttt{image\_normalised} = \frac{\texttt{image\_original}-\texttt{mean}}{\texttt{std}}.
\end{equation}
In our numerical experiments, we set \texttt{mean} = 0.5 and \texttt{std} = 0.5 to put the image pixel values within [-1, 1].

\subsection{Results and Analysis}

After training models with different levels of replacement five times with different parameter initialisations with all three datasets, we gather the averaged performance results of the last iteration in Table~\ref{tab:res-avg}. Detailed graphs on how performance metrics change during the optimisation process for both training and test data sets can be found in the Appendix~\ref{sec:Appendix-TorchPlots}. 

In general, the \texttt{HybridModel} with \texttt{replacement\_level = 2} achieves the best average performance in all three datasets. The average performance of all three different replacement levels degrades when the complexity of the data increases, from MNIST to FashionMNIST and then to CIFAR-10. Such performance degradation is likely due to the lack of expressivity of the network structure itself (normally in order to reach high performance on the CIFAR-10 dataset, one would need to follow the structure of neural networks like ResNet-18 \cite{ResNet}). It is also should be noted that operations such as Batch Normalisation and Layer Normalisation also play an important part in the performance of a neural network). Compared to previous results from the quantum machine learning literature on similar sets of data, such as in \cite{wang2024quantumhamiltonianembeddingimages}, our classical-quantum hybrid neural network achieves a higher performance compared to an end-to-end quantum neural network (96.9\% versus 89.7\% test accuracy on MNIST, and 86.6\% versus 79.6\% on FashionMNIST).

\begin{table}[ht!]
\caption{\label{tab:res-avg} Average of performance (loss and accuracy) results on all three datasets and with three levels of replacement. When the replacement level is 0, all layers in the neural network are classical, i.e. classical \texttt{Conv2d} and classical \texttt{Linear}; when the replacement level = 1, only the classical convolution layers are replaced with the quantum \texttt{FlippedQuanv3x3} layer; when the replacement level = 2, both the classical convlution and linear layers are replced with their quantum counterparts, i.e. \texttt{Conv2d}$\rightarrow$ \texttt{FlippedQuanv3x3} and \texttt{Linear}$\rightarrow$\texttt{DataReUploadingLinear}. We can see that the \texttt{DataReUploadingLinear} brought a non-trivial performance increase from replacement level = 1 to replacement level = 2.}
\tabcolsep=10pt\relax
\begin{tabular}{lcccc}
\toprule
\multirow{2}{*}{\textbf{Dataset}}&
\multirow{2}{*}{\textbf{Metric (Average Value)}}& \multicolumn{3}{c}{\textbf{Replacement Level}} \\
&&\textbf{\texttt{0}}&
\textbf{\texttt{1}}&
\textbf{\texttt{2}}\\
\midrule
\multirow{4}{*}{MNIST}&\textrm{Train Loss} & 0.229 & 0.227 & \textbf{0.076}\\
&\textrm{Test Loss} & 0.294 & 0.295 & \textbf{0.125} \\
&\textrm{Train Accuracy} & 93.5\% & 93.6\% & \textbf{97.8\%}\\
&\textrm{Test Accuracy} & 92.3\% & 92.3\%  & \textbf{96.9\%}\\
\midrule
\multirow{4}{*}{FashionMNIST}&\textrm{Train Loss} & 0.350 & 0.353 & \textbf{0.280} \\
&\textrm{Test Loss} & 0.476 & 0.477 & \textbf{0.405} \\
&\textrm{Train Accuracy} & 87.6\% & 87.5\% & \textbf{90.1\%} \\
&\textrm{Test Accuracy} & 83.6\% & 83.4\%  & \textbf{86.6\%}\\
\midrule
\multirow{4}{*}{CIFAR-10}&\textrm{Train Loss} & 1.50 & 1.61 & \textbf{1.28}\\
&\textrm{Test Loss} & 1.89 & 1.78 & \textbf{1.49} \\
&\textrm{Train Accuracy} & 48.9\% & 45.0\% & \textbf{55.0\%}\\
&\textrm{Test Accuracy} & 37.0\% & 38.9\%  & \textbf{48.9\%}\\
\botrule
\end{tabular}
\end{table}

\begin{table}[ht!]
\caption{\label{tab:res-train-test-diff} The performance difference between training and testing on all three datasets and with three levels of replacement, which highlights the model's capability to generalise from training data to test data. When replacement level is 0, all the layers in the neural network are classical, i.e. classical \texttt{Conv2d} and classical \texttt{Linear}; when replacement level = 1, only the classical convolution layers are replaced with the quantum \texttt{FlippedQuanv3x3} layer; when replacement level = 2, both the classical convlution and linear layers are replced with their quantum counterparts, i.e. \texttt{Conv2d}$\rightarrow$ \texttt{FlippedQuanv3x3} and \texttt{Linear}$\rightarrow$\texttt{DataReUploadingLinear}.}
\tabcolsep=5pt\relax
\begin{tabular}{lcccc}
\toprule
\multirow{3}{*}{\textbf{Dataset}}&
\multirow{3}{*}{\textbf{Metric}} & \multicolumn{3}{c}{\texttt{test\_metric-train\_metric}} \\ && \multicolumn{3}{c}{\textbf{Replacement Level}  } \\
&&\textbf{\texttt{0}}&
\textbf{\texttt{1}}&
\textbf{\texttt{2}}\\
\midrule
\multirow{2}{*}{MNIST}&\textrm{Loss}& 0.0648 & 0.0685 & 0.0487\\

&\textrm{Accuracy} & -0.0121 & -0.0126 & -0.00905 \\

\midrule
\multirow{2}{*}{FashionMNIST}&\textrm{Loss} & 0.125 & 0.124 & 0.125 \\

&\textrm{Accuracy} & -0.0393 & -0.0409 & -0.0354 \\

\midrule
\multirow{2}{*}{CIFAR-10}&\textrm{Loss} & 0.392 & 0.166 & 0.202 \\

&\textrm{Accuracy} & -0.120 & -0.0614 & -0.0610 \\

\botrule
\end{tabular}
\end{table}

\begin{table}[ht!]
\caption{\label{tab:res-std} The Standard deviation of performance (loss and accuracy) results on all three datasets and with three levels of replacement. When replacement level is 0, all the layers in the neural network are classical, i.e. classical \texttt{Conv2d} and classical \texttt{Linear}; when replacement level = 1, only the classical convolution layers are replaced with the quantum \texttt{FlippedQuanv3x3} layer; when replacement level = 2, both the classical convolution and linear layers are replced with their quantum counterparts, i.e. \texttt{Conv2d}$\rightarrow$ \texttt{FlippedQuanv3x3} and \texttt{Linear}$\rightarrow$\texttt{DataReUploadingLinear}.}
\tabcolsep=5pt\relax
\begin{tabular}{lcccc}
\toprule
\multirow{4}{*}{\textbf{Dataset}}&
\multirow{4}{*}{\textbf{Metric} } &  \multicolumn{3}{c}{\textbf{Standard Deviation}}\\&& \multicolumn{3}{c}{ ($\times 10^{-3}$) } \\ && \multicolumn{3}{c}{\textbf{Replacement Level}} \\
& &\textbf{\texttt{0}}&
\textbf{\texttt{1}}&
\textbf{\texttt{2}}\\
\midrule
\multirow{4}{*}{MNIST}&\textrm{Train Loss}& 0.410 & 0.425 & 5.65\\
&\textrm{Test Loss} & 1.07 & 4.38 & 15.9 \\
&\textrm{Train Accuracy} & 0.173 & 0.413 & 1.30\\
&\textrm{Test Accuracy}  & 1.33 & 1.29  & 3.24\\
\midrule
\multirow{4}{*}{FashionMNIST}&\textrm{Train Loss} & 0.850 & 0.368 & 3.56 \\
&\textrm{Test Loss} & 1.61 & 7.31 & 19.2 \\
&\textrm{Train Accuracy} & 0.571 & 0.216 & 0.680 \\
&\textrm{Test Accuracy} & 1.38 & 3.69  & 6.16\\
\midrule
\multirow{4}{*}{CIFAR-10}&\textrm{Train Loss} & 1.38 & 4.94 & 11.8 \\
&\textrm{Test Loss} & 2.11 & 2.26 & 12.5 \\
&\textrm{Train Accuracy} & 0.978 & 1.33 & 4.93 \\
&\textrm{Test Accuracy} & 3.06 & 2.58  & 0.866\\
\botrule
\end{tabular}
\end{table}

We also notice that, simply replacing the classical \texttt{Conv2d} layer with \texttt{FlippedQuanv3x3} layer does not bring a significant performance change on all three datasets. This can be explained through the analysis in Section~\ref{sec:flipped-quanv}, which shows that mathematically the \texttt{FlippedQuanv3x3} layer has little difference from the \texttt{Conv2d} layer. Major performance boost occurs when the classical \texttt{Linear} layer is replaced with the quantum \texttt{DataReUploadingLinear}. In \cite{wang2024quantumhamiltonianembeddingimages}, it was shown that the transformation applied to input data ($H_M$) by the \texttt{DataReUploadingLinear} is nonlinear:
\begin{equation}\label{eqn:emb-expand}
    W(H_M; t=\frac{1}{L}) = 1-\frac{{i}H_M}{2^1 L}-\frac{H_M^2}{2!\times 2^2 L^2}+\frac{{i} H_M^3}{3!\times 2^3 L^3}+ \cdots,
\end{equation}
where we set $t = \frac{1}{L}$, $H_M = \frac{M+M^T}{2}$, $M$ is the padded, reshaped feature map from the previous layer, and $L$ is the number of layers in the data reuploading circuit in the \texttt{DataReUploadingLinear} layer. We can see that the nonlinearity provided by Eqn.~\ref{eqn:emb-expand} is much stronger than the classical linear layer, which is just the affine transformation:
\begin{equation}
    \texttt{Linear}(\boldsymbol{x}) = \boldsymbol{W}^T \boldsymbol{x}+\boldsymbol{b},
\end{equation}
where $\boldsymbol{x}$ is the input, $\boldsymbol{W}$ is the weight matrix and $\boldsymbol{b}$ is the bias term. This indicates that combining the \verb+FlippedQuanv3x3+ with quantum Hamiltonian embedding could potentially outperform classical convolution operation without the need for activation functions such ReLU. From Table~\ref{tab:res-train-test-diff}, which highlights the model's capability to generalise from training data to test data, we could also see that, compared to the fully classical model, quantum models have a slight advantage on generalisation.

A major problem of quantum-based approaches is the optimisation (loss) landscape. From the standard deviation of the performance metrics (Table~\ref{tab:res-std}), we can see that generally the performance of the quantum approaches is more sensitive to parameter initialisation than classical models, especially when \texttt{replacement\_level} = 2. This could be due to the fact that the parameters in a quantum model are not ``directly" applied to the input, like classical models. The optimisation landscape of quantum models is further distorted by the implicit trigonometric functions within parameterised quantum gates compared to classical models. This could present further challenges on the training of quantum models.

\section{\label{sec:discussion}Discussion}


In this paper, we proposed a scheme for developing quantum neural network models and comparing their performance with their classical counterparts by gradually swapping classical components of the classical layers and keeping the information-passing structure, instead of designing an end-to-end quantum neural network model. During the development of classical deep learning models, the information-passing structure is as important as the design of individual layers, if not more important. For example, the major difference between ResNet \cite{ResNet} and ``plain" convolutional neural networks like VGG \cite{VGG} is the information passing structure rather than the convolution and linear layers themselves. Classical information-passing structure allows more diverse operations than a fully quantum structure, such as the copy and concatenation of feature maps. Hence, it would be preferable to preserve the information-passing structure of classical neural networks and only replace individual layers with their quantum counterparts.

To replace the classical \texttt{Conv2d} layer, we propose the \texttt{FlippedQuanv3x3} layer based on the flipped model of quantum machine learning \cite{Jerbi2023-rg}. To replace the classical \texttt{Linear} layer, we propose the \verb+DataReUploadingLinear+ layer based on the quantum Hamiltonian embedding proposed in \cite{wang2024quantumhamiltonianembeddingimages}. Since mathematically speaking, the \texttt{FlippedQuanv3x3} layer does very similar things as the classical convolution operation, only swapping \texttt{Conv2d} with \texttt{FlippedQuanv3x3} does not change the performance by a large margin. However, replacing the classical linear layer with the \verb+DataReUploadingLinear+ layer has a positive influence on the averaged performance on all three datasets. The reason for this could originate from the implicit nonlinear transformation of the input brought by the quantum Hamiltonian embedding. Also, replacing classical layers with their quantum counterpart increases the model's sensitivity to parameter initialisation.

From the data collected in Tables~\ref{tab:res-avg}, \ref{tab:res-std} and \ref{tab:res-train-test-diff}, we can clearly see the trend of performance change when we gradually introduce more quantum components into the neural network architecture. This process provides us with a more fine-grained characterization of the advantage/disadvantage brought by quantum computing to the neural network model, compared to other benchmarking approaches which directly compare an end-to-end quantum model with classical models, such as those presented \cite{Bowles2024-kb}. 

In general, our framework provides a new approach for designing quantum neural network models as well as benchmarking these models against their classical counterparts. Our results show that, when given the same classical information-passing structure, neural networks with quantum layers could achieve a similar, if not better, level of performance compared to NNs with only classical layers. Although it is sill too early to claim any sort of quantum advantage for QNN, our research highlights the need for further investigation on the potential of QNN. It is also worth mentioning that the design philosophy of our approach -- building quantum models by composing smaller components with a certain structure instead of an end-to-end monolithic quantum model, coincides with the compositional approach for designing intelligent systems, which has been advocated by researchers from classical AI \cite{Du2024-cw, GuptaUnknown-vf}, quantum AI \cite{Tull2024-th} and even cognitive science \cite{Friston2024-nx}.

\backmatter





\bmhead{Acknowledgements}

This research was supported by the University of Melbourne through the establishment of the IBM Quantum Network Hub at the University.

\clearpage
\begin{appendices}

\section{\label{sec:Appendix-TorchPlots} Performance Plots During Training and Testing}

\begin{figure}[ht!]
    \centering
    \includegraphics{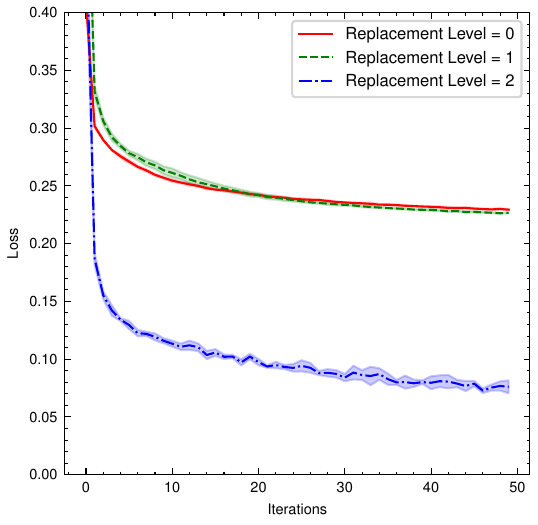}
    \caption{Loss plot during the training on the MNIST dataset for different replacement levels. The Y-axis is cut off at 0.40 for clearer presentation of the loss differences towards the end of the training. }
    \label{fig:torch-mnist-train-loss}
\end{figure}

\begin{figure}[ht!]
    \centering
    \includegraphics{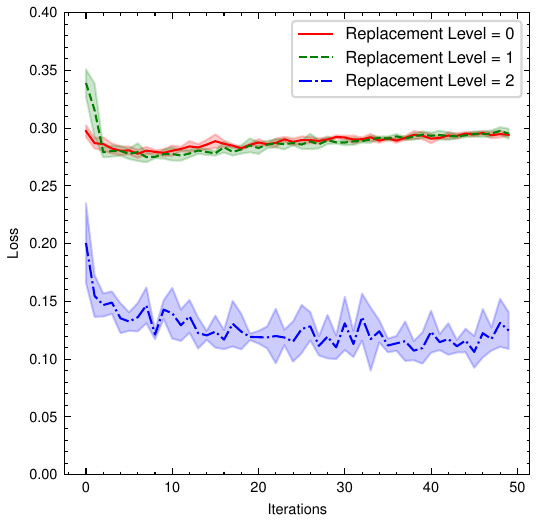}
    \caption{Loss plot during the testing on the MNIST dataset for different replacement levels. The Y-axis is cut off at 0.40 for clearer presentation of the loss differences towards the end of the 50 iterations.}
    \label{fig:torch-mnist-test-loss}
\end{figure}

\begin{figure}[ht!]
    \centering
    \includegraphics{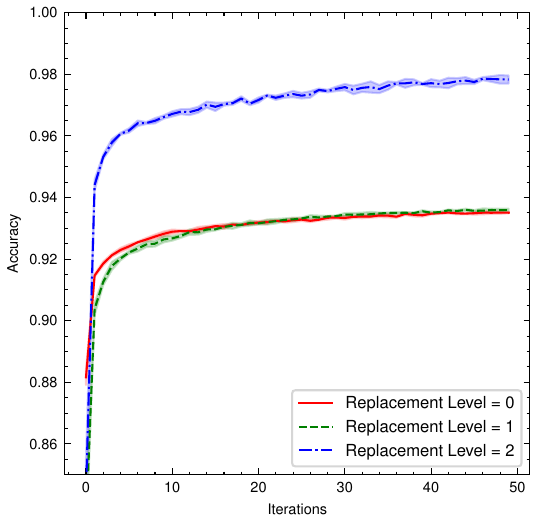}
    \caption{Accuracy plot during the training on the MNIST dataset for different replacement levels. The Y-axis is cut off between 0.85 and 1.00 for clearer presentation of the accuracy differences towards the end of the 50 iterations.}
    \label{fig:torch-mnist-train-acc}
\end{figure}

\begin{figure}[ht!]
    \centering
    \includegraphics{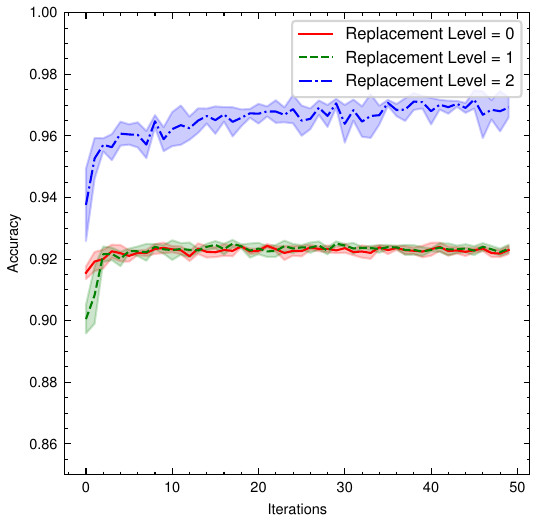}
    \caption{Accuracy plot during the testing stage on the MNIST dataset for different replacement levels. The Y-axis is cut off between 0.85 and 1.00 for clearer presentation of the accuracy differences towards the end of the 50 iterations.}
    \label{fig:torch-mnist-test-acc}
\end{figure}

\begin{figure}[ht!]
    \centering
    \includegraphics{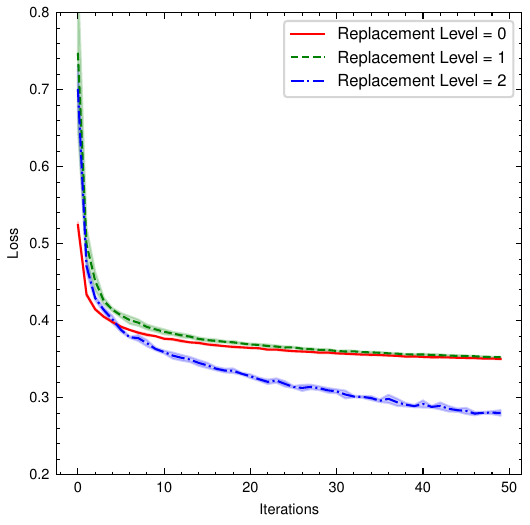}
    \caption{Loss plot during the training on the FashionMNIST dataset for different replacement levels. The Y-axis is cut off between 0.2 and 0.8 for clearer presentation of the loss differences towards the end of the training. }
    \label{fig:torch-fashionmnist-train-loss}
\end{figure}

\begin{figure}[ht!]
    \centering
    \includegraphics{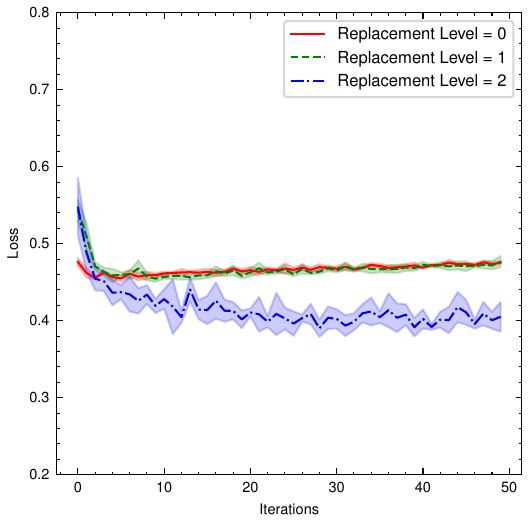}
    \caption{Loss plot during the testing stage on the FashionMNIST dataset for different replacement levels. The Y-axis is cut off between 0.2 and 0.8 for clearer presentation of the loss differences towards the end of the training. }
    \label{fig:torch-fashionmnist-test-loss}
\end{figure}

\begin{figure}[ht!]
    \centering
    \includegraphics{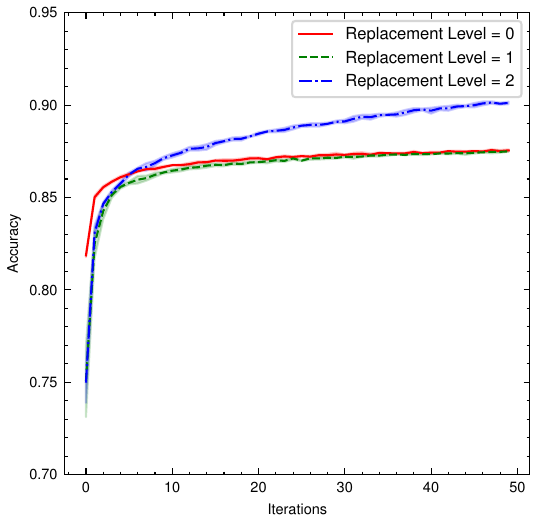}
    \caption{Accuracy plot during the training on the FashionMNIST dataset for different replacement levels. The Y-axis is cut off between 0.7 and 0.95 for clearer presentation of the accuracy differences towards the end of the training. }
    \label{fig:torch-fashionmnist-train-acc}
\end{figure}

\begin{figure}[ht!]
    \centering
    \includegraphics{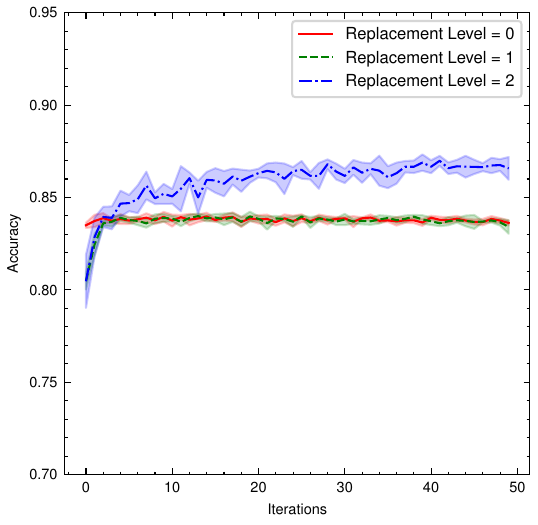}
    \caption{Accuracy plot during the testing stage on the FashionMNIST dataset for different replacement levels. The Y-axis is cut off between 0.7 and 0.95 for clearer presentation of the accuracy differences towards the end of the training. }
    \label{fig:torch-fashionmnist-test-acc}
\end{figure}

\begin{figure}[ht!]
    \centering
    \includegraphics{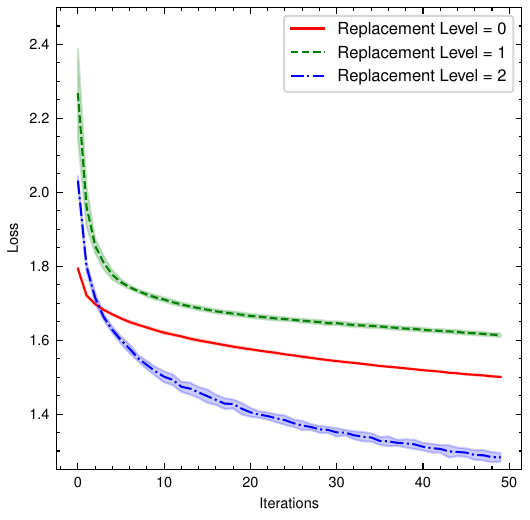}
    \caption{Loss plot during the training on the CIFAR-10 dataset for different replacement levels. The Y-axis is cut off between 1.25 and 2.5 for clearer presentation of the loss differences towards the end of the training. }
    \label{fig:torch-cifar10-train-loss}
\end{figure}

\begin{figure}[ht!]
    \centering
    \includegraphics{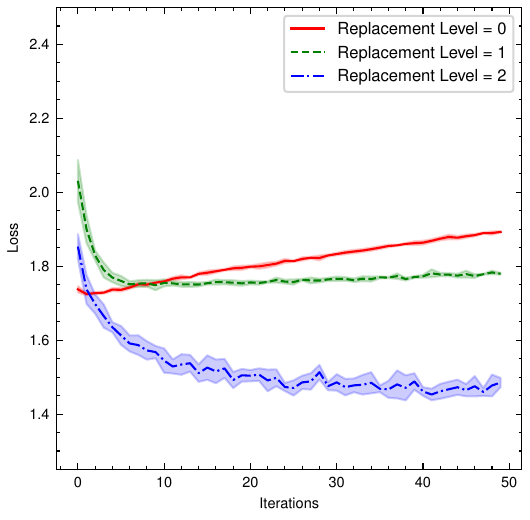}
    \caption{Loss plot during the testing stage on the FashionMNIST dataset for different replacement levels. The Y-axis is cut off between 1.25 and 2.5 for clearer presentation of the loss differences towards the end of the training. }
    \label{fig:torch-cifar10-test-loss}
\end{figure}

\begin{figure}[ht!]
    \centering
    \includegraphics{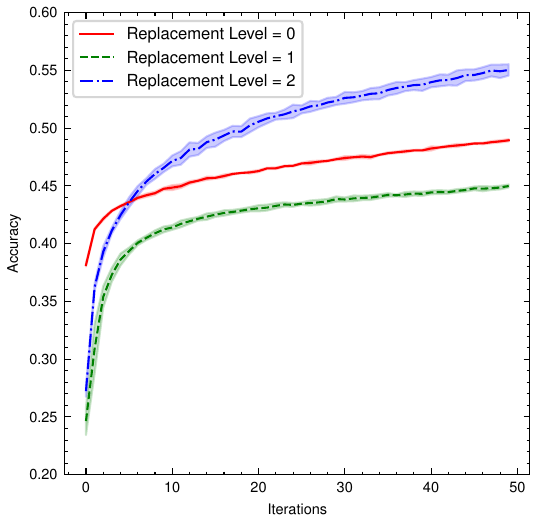}
    \caption{Accuracy plot during the training on the CIFAR-10 dataset for different replacement levels. The Y-axis is cut off between 0.2 and 0.6 for clearer presentation of the accuracy differences towards the end of the training. }
    \label{fig:torch-cifar10-train-acc}
\end{figure}

\begin{figure}[ht!]
    \centering
    \includegraphics{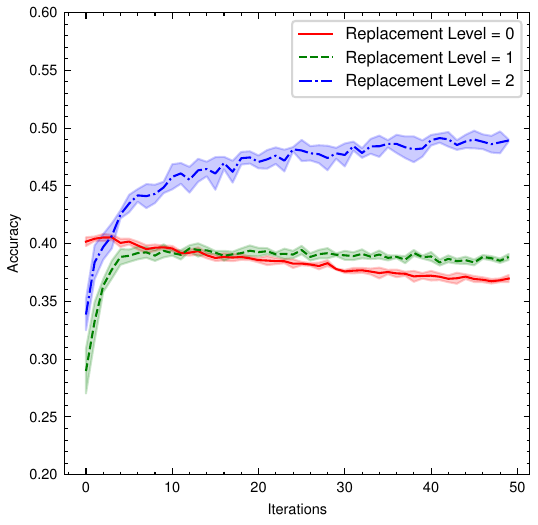}
    \caption{Accuracy plot during the testing stage on the CIFAR-10 dataset for different replacement levels. The Y-axis is cut off between 0.2 and 0.6 for clearer presentation of the accuracy differences towards the end of the training. }
    \label{fig:torch-cifar10-test-acc}
\end{figure}

The average performance (loss and accuracy) plots during the training and testing iterations are shown from Fig.~\ref{fig:torch-mnist-train-loss} to Fig.~\ref{fig:torch-cifar10-test-acc}. The performance curves are shaded, with the lower bound of the shade equals to \texttt{avg-std} and the upper bound of the shade equals to \texttt{avg+std}, where \texttt{avg} is the averaged performance metrics (loss and accuracy) and \texttt{std} is the standard deviation of the performance metrics, both over five repetitions.

\clearpage
\section{\label{sec:Appendix-JAX}Impact of Hybrid Model Performance Due To Simulation Software}

\subsection{Overview of the JAX-Simulated Results}


We also implemented the simulation of the quantum model using JAX \cite{jax2018github} and the JAX-based  neural network library \verb+Equinox+ \cite{kidger2021equinox}. The classical optimiser used is from Optax \cite{deepmind2020jax}. The data preprocess methods are the same as the PyTorch-based simulations. The performance of the JAX-simulated models is different from the PyTorch-simulated models in the main text. 

From Table~\ref{tab:res-avg-jax}, we could see that the loss values are much larger than those achieved by the PyTorch-based experiments in the main text, even though both are performed in FP32 precision\footnote{Increasing the precision from FP32 to FP64 would not change the performance by a noticable margin, but would increase the memory as well as time required for the simulation.}. On the MNIST dataset, we can see that the accuracy for \texttt{replacement\_level}=2 is not far from \texttt{replacement\_level}=0, however the loss is much larger, indicating that the output classification scores are more evenly distributed among different labels. 

In Table~\ref{tab:res-std-jax}, we could see that the standard deviation is large even when \texttt{replacement\_level}=0, compared to the standard deviation of the PyTorch-based simulation shown in Table~\ref{tab:res-std} in the main text, even with a larger batch size. However, the standard deviation of the JAX-simulated models with \texttt{replacement\_level}=3 is smaller than that of the PyTorch-simulated models on average. This could be attributed to the difference in the implementation details of the Adam optimiser between \verb+torch.optim.Adam+ and \verb+optax.adam+, as well as the numerical details between the linear algebra operations in PyTorch and those in the JIT-compiled JAX code.

Also, as shown in Table~\ref{tab:res-train-test-diff-jax}, the performance gap on train and test datasets of JAX-simulated models is smaller than that of the PyTorch-based simulations. Again, this could be attributed to the differences in the implementation of both the Adam optimiser and the linear algebra operations.

\begin{table}[ht!]
\caption{\label{tab:hyperparam-jax} Training hyperparameters. The rest of the hyperparameters in the classical optimiser are kept the same as the default setting. The simulation is performed on a NVIDIA A100-40G GPU on Google Colab. For each \texttt{replacement\_level}-dataset combination, the training is repeated five times with different paraemter initialisations.}
\begin{tabular}{cccc}
\toprule
\textbf{Batch Size} & \textbf{Initial Learning Rate} & \textbf{Number of Iterations} &  \textbf{Optimiser} \\
\midrule
1000 & $5\times 10^{-3}$ & 100 & \verb+optax.adam+\\
\botrule
\end{tabular}
\end{table}

\begin{table}[ht!]
\caption{\label{tab:res-avg-jax} Average of performance (loss and accuracy) results on all three datasets and with three levels of replacement for JAX-based simulations. When replacement level is 0, all the layers in the neural network are classical, i.e. classical \texttt{Conv2d} and classical \texttt{Linear}; when replacement level = 1, only the classical convolution layers are replaced with the quantum \texttt{FlippedQuanv3x3} layer; when replacement level = 2, both the classical convlution and linear layers are replced with their quantum counterparts, i.e. \texttt{Conv2d}$\rightarrow$ \texttt{FlippedQuanv3x3} and \texttt{Linear}$\rightarrow$\texttt{DataReUploadingLinear}.}
\tabcolsep=10pt\relax
\begin{tabular}{lcccc}
\toprule
\multirow{2}{*}{\textbf{Dataset}}&
\multirow{2}{*}{\textbf{Metric (Average Value)}}& \multicolumn{3}{c}{\textbf{Replacement Level}} \\
&&\textbf{\texttt{0}}&
\textbf{\texttt{1}}&
\textbf{\texttt{2}}\\
\midrule
\multirow{4}{*}{MNIST}&\textrm{Train Loss} &0.235  &0.234   & 1.70 \\
&\textrm{Test Loss} & 0.300  & 0.290  & 1.69  \\
&\textrm{Train Accuracy} & 93.6\%  & 93.6\%  & 94.5\% \\
&\textrm{Test Accuracy} & 92.1\%  & 92.2\%   & 94.5\% \\
\midrule
\multirow{4}{*}{FashionMNIST}&\textrm{Train Loss} & 0.356  & 0.356  & 1.75  \\
&\textrm{Test Loss} & 0.477 &  0.473 & 1.76  \\
&\textrm{Train Accuracy} &  87.6\% & 87.6\%  &  81.8\% \\
&\textrm{Test Accuracy} & 83.7\%  & 83.7\%   & 80.5\% \\
\midrule
\multirow{4}{*}{CIFAR-10}&\textrm{Train Loss} & 1.49  & 1.63  & 2.15 \\
&\textrm{Test Loss} & 2.06  & 1.78  &  2.15 \\
&\textrm{Train Accuracy} & 50.6\%  & 44.8\%  & 39.1\% \\
&\textrm{Test Accuracy} &  33.7\% &  38.7\% & 38.1\% \\
\botrule
\end{tabular}
\end{table}

\begin{table}[ht!]
\caption{\label{tab:res-std-jax} The standard deviation of performance (loss and accuracy) results on all three datasets and with three levels of replacement, models simulated with JAX. When replacement level is 0, all the layers in the neural network are classical, i.e. classical \texttt{Conv2d} and classical \texttt{Linear}; when replacement level = 1, only the classical convolution layers are replaced with the quantum \texttt{FlippedQuanv3x3} layer; when replacement level = 2, both the classical convlution and linear layers are replced with their quantum counterparts, i.e. \texttt{Conv2d}$\rightarrow$ \texttt{FlippedQuanv3x3} and \texttt{Linear}$\rightarrow$\texttt{DataReUploadingLinear}.}
\tabcolsep=5pt\relax
\begin{tabular}{lcccc}
\toprule
\multirow{4}{*}{\textbf{Dataset}}&
\multirow{4}{*}{\textbf{Metric} } &  \multicolumn{3}{c}{\textbf{Standard Deviation}}\\&& \multicolumn{3}{c}{ ($\times 10^{-3}$) } \\ && \multicolumn{3}{c}{\textbf{Replacement Level}} \\
&&\textbf{\texttt{0}}&
\textbf{\texttt{1}}&
\textbf{\texttt{2}}\\
\midrule
\multirow{4}{*}{MNIST}&\textrm{Train Loss}& 2.16   & 1.49  & 0.313 \\
&\textrm{Test Loss} & 4.69  & 3.87  & 0.886  \\
&\textrm{Train Accuracy} &  0.533 & 0.878  & 0.261 \\
&\textrm{Test Accuracy}  &  1.52 &   0.960 & 1.10 \\
\midrule
\multirow{4}{*}{FashionMNIST}&\textrm{Train Loss} & 2.61  & 1.71  &  0.984 \\
&\textrm{Test Loss} &  9.45 & 3.00  &  0.710 \\
&\textrm{Train Accuracy} & 0.669  &  0.722 &  0.616 \\
&\textrm{Test Accuracy} &  3.18 &   2.76 & 0.049 \\
\midrule
\multirow{4}{*}{CIFAR-10}&\textrm{Train Loss} & 6.55  &  5.81 & 3.97 \\
&\textrm{Test Loss} & 31.7  & 7.99  &  3.73 \\
&\textrm{Train Accuracy} &  2.69 & 2.34  & 10.2  \\
&\textrm{Test Accuracy} & 6.95 &   3.45  & 9.86 \\
\botrule
\end{tabular}
\end{table}

\begin{table}[ht!]
\caption{\label{tab:res-train-test-diff-jax} The performance difference between training and testing on all three datasets and with three levels of replacement, which highlights the model's capability to generalise from training data to test data, models simulated with JAX. When replacement level is 0, all the layers in the neural network are classical, i.e. classical \texttt{Conv2d} and classical \texttt{Linear}; when replacement level = 1, only the classical convolution layers are replaced with the quantum \texttt{FlippedQuanv3x3} layer; when replacement level = 2, both the classical convolution and linear layers are replced with their quantum counterparts, i.e. \texttt{Conv2d}$\rightarrow$ \texttt{FlippedQuanv3x3} and \texttt{Linear}$\rightarrow$\texttt{DataReUploadingLinear}.}
\tabcolsep=5pt\relax
\begin{tabular}{lcccc}
\toprule
\multirow{3}{*}{\textbf{Dataset}}&
\multirow{3}{*}{\textbf{Metric}} & \multicolumn{3}{c}{\texttt{test\_metric-train\_metric}} \\ && \multicolumn{3}{c}{\textbf{Replacement Level}  } \\
&&\textbf{\texttt{0}}&
\textbf{\texttt{1}}&
\textbf{\texttt{2}}\\
\midrule
\multirow{2}{*}{MNIST}&\textrm{Loss}& 0.0647  & 0.0563  &  -0.00410\\

&\textrm{Accuracy} &  -0.0157 &  -0.0137 &  0.000260 \\

\midrule
\multirow{2}{*}{FashionMNIST}&\textrm{Loss} &  0.121 & 0.116  & 0.00767  \\

&\textrm{Accuracy} & -0.0393  & -0.0393  &  -0.0122 \\

\midrule
\multirow{2}{*}{CIFAR-10}&\textrm{Loss} &  0.567 &  0.151 &  0.000373 \\

&\textrm{Accuracy} &  -0.169 & -0.0617  &  -0.0106 \\

\botrule
\end{tabular}
\end{table}

Since both the JAX-simulated and PyTorch-simulated models are trained with FP32 prescision, and the linear algebra functions such as \texttt{torch.matrix\_exp} and \texttt{jax.scipy.linalg.expm}, and the numerical difference for the same input is on the scale of $10^{-7}$, which cannot explain the large discrepancies (at least on the scale of $10^{-2}$) between the results from the two simulation backend. We conjecture that the main reason lies in initialisation, so we calculate the performance results in the first iteration of training and show them in Tables~\ref{tab:res-avg-torch-first-iter} and ~\ref{tab:res-avg-jax-first-iter}.

\begin{table}[ht!]
\caption{\label{tab:res-avg-torch-first-iter} Average of performance (loss and accuracy) results on all three datasets and with three levels of replacement for PyTorch-based simulations on the first iteration. When replacement level is 0, all the layers in the neural network are classical, i.e. classical \texttt{Conv2d} and classical \texttt{Linear}; when replacement level = 1, only the classical convolution layers are replaced with the quantum \texttt{FlippedQuanv3x3} layer; when replacement level = 2, both the classical convlution and linear layers are replced with their quantum counterparts, i.e. \texttt{Conv2d}$\rightarrow$ \texttt{FlippedQuanv3x3} and \texttt{Linear}$\rightarrow$\texttt{DataReUploadingLinear}.}
\begin{tabular}{lcccc}
\toprule
\multirow{2}{*}{\textbf{Dataset}}&
\multirow{2}{*}{\textbf{Metric (Average Value, 1st Iteration)}}& \multicolumn{3}{c}{\textbf{Replacement Level}} \\
&&\textbf{\texttt{0}}&
\textbf{\texttt{1}}&
\textbf{\texttt{2}}\\
\midrule
\multirow{4}{*}{MNIST}&\textrm{Train Loss} & 0.410  & 0.583   & 0.486 \\
&\textrm{Test Loss} & 0.297  & 0.339  & 0.200  \\
&\textrm{Train Accuracy} & 88.2\%  & 83.3\%  & 84.3\% \\
&\textrm{Test Accuracy} & 91.5\%  & 90.5\%   & 93.7\% \\
\midrule
\multirow{4}{*}{FashionMNIST}&\textrm{Train Loss} & 0.524  & 0.747  & 0.701  \\
&\textrm{Test Loss} & 0.476 &  0.543 & 0.548  \\
&\textrm{Train Accuracy} &  81.9\% &  75.1\%  &  75.0\% \\
&\textrm{Test Accuracy} & 83.5\%  & 80.5\%   & 80.5\% \\
\midrule
\multirow{4}{*}{CIFAR-10}&\textrm{Train Loss} & 1.79  & 2.27  & 2.03 \\
&\textrm{Test Loss} & 1.74  & 2.03  &  1.85 \\
&\textrm{Train Accuracy} & 38.2\%  & 24.6\%  & 27.2\% \\
&\textrm{Test Accuracy} &  40.2\% &  29.0\% & 33.9\% \\
\botrule
\end{tabular}
\end{table}

\begin{table}[ht!]
\caption{\label{tab:res-avg-jax-first-iter} Average of performance (loss and accuracy) results on all three datasets and with three levels of replacement for JAX and Equinox-based simulations on the first iteration. When replacement level is 0, all the layers in the neural network are classical, i.e. classical \texttt{Conv2d} and classical \texttt{Linear}; when replacement level = 1, only the classical convolution layers are replaced with the quantum \texttt{FlippedQuanv3x3} layer; when replacement level = 2, both the classical convlution and linear layers are replced with their quantum counterparts, i.e. \texttt{Conv2d}$\rightarrow$ \texttt{FlippedQuanv3x3} and \texttt{Linear}$\rightarrow$\texttt{DataReUploadingLinear}.}
\begin{tabular}{lcccc}
\toprule
\multirow{2}{*}{\textbf{Dataset}}&
\multirow{2}{*}{\textbf{Metric (Average Value, 1st Iteration)}}& \multicolumn{3}{c}{\textbf{Replacement Level}} \\
&&\textbf{\texttt{0}}&
\textbf{\texttt{1}}&
\textbf{\texttt{2}}\\
\midrule
\multirow{4}{*}{MNIST}&\textrm{Train Loss} & 2.14  &  18.4   & 2.42 \\
&\textrm{Test Loss} & 0.342  & 0.752  & 2.16  \\
&\textrm{Train Accuracy} & 77.4\%  & 48.6\%  & 11.8\% \\
&\textrm{Test Accuracy} & 90.3\%  & 82.9\%   & 16.3\% \\
\midrule
\multirow{4}{*}{FashionMNIST}&\textrm{Train Loss} & 1.60  & 12.0  & 2.35  \\
&\textrm{Test Loss} & 0.523 &  1.21 & 2.12  \\
&\textrm{Train Accuracy} &  72.5\% &  56.8\%  &  11.6\% \\
&\textrm{Test Accuracy} & 81.9\%  & 75.2\%   & 16.2\% \\
\midrule
\multirow{4}{*}{CIFAR-10}&\textrm{Train Loss} & 2.04  & 12.0 & 2.50 \\
&\textrm{Test Loss} & 1.75  & 2.03  &  2.38 \\
&\textrm{Train Accuracy} & 35.7\%  & 20.8\%  & 10.3\% \\
&\textrm{Test Accuracy} &  39.8\% &  28.0\% & 10.7\% \\
\botrule
\end{tabular}
\end{table}

Comparing the performance results of the first iteration of training and testing shown in Table~\ref{tab:res-avg-torch-first-iter} and Table~\ref{tab:res-avg-jax-first-iter}, we can see that the performance of the JAX and Equinox-simulated models is usually worse than that of PyTorch-simulated ones. The performance gap between the PyTorch-simulated and JAX-simulated models widens when \texttt{replacement\_level} changes from \texttt{0} to \texttt{2}, especially when \texttt{replacement\_level = 2}. In the implementation, the parameter initialisation of the classical layers (\texttt{Conv2d} and \texttt{Linear}) is handled internally by the neural network libraries (PyTorch and Equinox). However, for the (simulated) quantum layers, parameter initialisation must be implemented by hand. For PyTorch-simulated models, parameters are initialised with \texttt{torch.randn} without specifying any random seeds; For JAX and Equinox-simulated models, parameters are initialsed with \texttt{jax.random.normal}, which requires a \texttt{jax.random.PRNGKey} as input. The random key is generated from a single seed, and split via \texttt{jax.random.split}. This difference could have caused the difference in the averaged performance of the initial iteration during training. With such a performance gap at the beginning, it could lead to the performance gap between PyTorch-based models and JAX/Equinox-based ones. For example, for the FashionMNIST dataset, the accuracy of the initial iteration of the PyTorch-simulated model when \texttt{replacement\_level = 2} in the test data is already close to the last iteration performance of the JAX-simulated model.

\subsection{\label{sec:Appendix-JAXPlots} Performance Plots During Training and Testing for JAX-based Experiments}

\begin{figure}[ht!]
    \centering
    \includegraphics{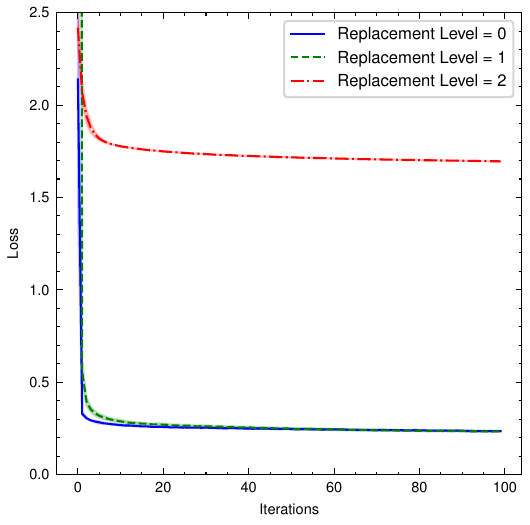}
    \caption{Loss plot during the training on the MNIST dataset for different replacement levels. The Y-axis is cut off at 2.5 for clearer presentation of the loss differences towards the end of the training. }
    \label{fig:eqx-mnist-train-loss}
\end{figure}

\begin{figure}[ht!]
    \centering
    \includegraphics{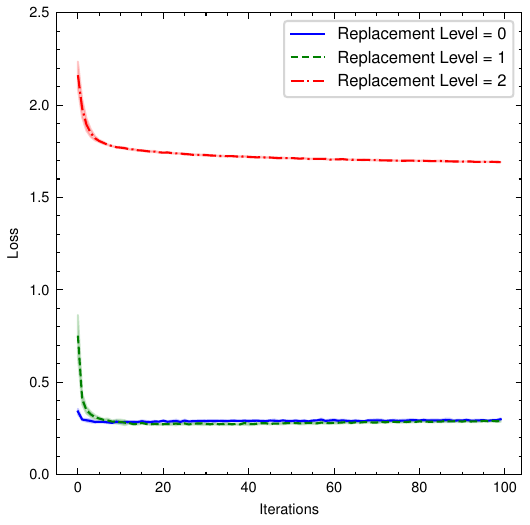}
    \caption{Loss plot during the testing on the MNIST dataset for different replacement levels. The Y-axis is cut off at 2.5 for clearer presentation of the loss differences towards the end of the 100 iterations.}
    \label{fig:eqx-mnist-test-loss}
\end{figure}

\begin{figure}[ht!]
    \centering
    \includegraphics{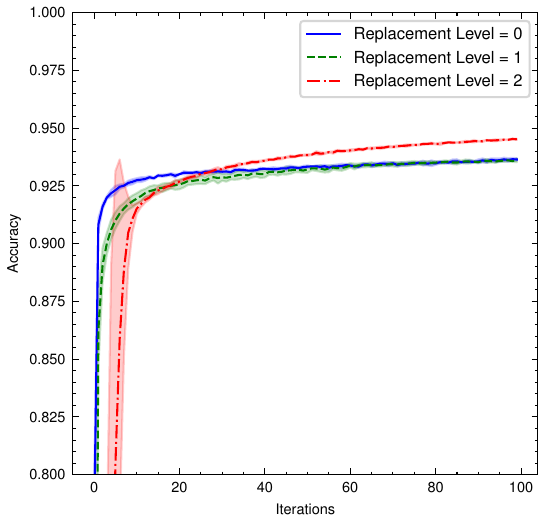}
    \caption{Accuracy plot during the training on the MNIST dataset for different replacement levels. The Y-axis is cut off between 0.8 and 1 for clearer presentation of the accuracy differences towards the end of the 100 iterations.}
    \label{fig:eqx-mnist-train-acc}
\end{figure}

\begin{figure}[ht!]
    \centering
    \includegraphics{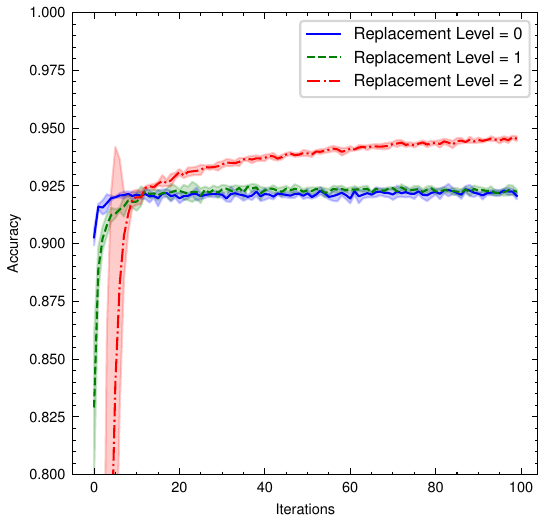}
    \caption{Accuracy plot during the testing stage on the MNIST dataset for different replacement levels. The Y-axis is cut off between 0.8 and 1 for clearer presentation of the accuracy differences towards the end of the 100 iterations.}
    \label{fig:eqx-mnist-test-acc}
\end{figure}

\begin{figure}[ht!]
    \centering
    \includegraphics{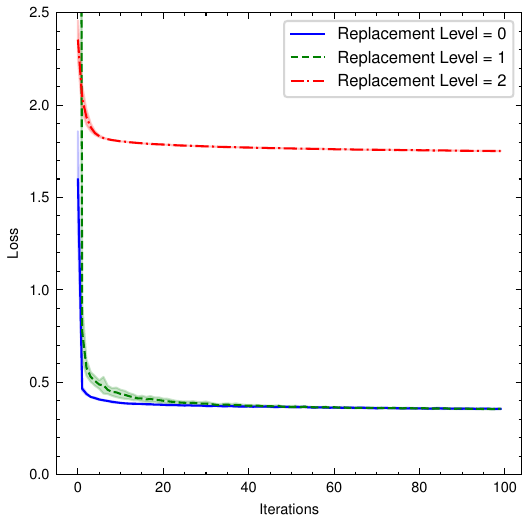}
    \caption{Loss plot during the training on the FashionMNIST dataset for different replacement levels. The Y-axis is cut off between 0 and 2.5 for clearer presentation of the loss differences towards the end of the training. }
    \label{fig:eqx-fashionmnist-train-loss}
\end{figure}

\begin{figure}[ht!]
    \centering
    \includegraphics{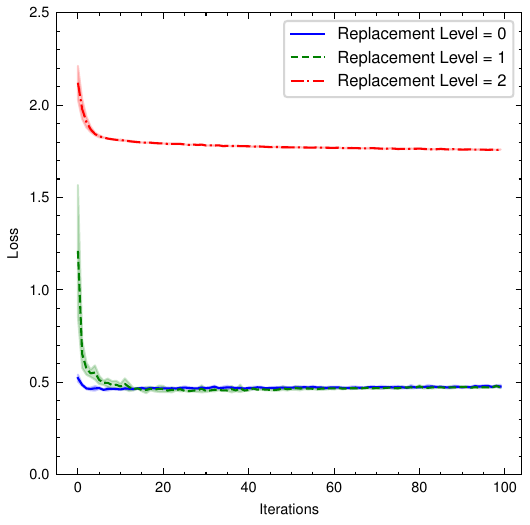}
    \caption{Loss plot during the testing stage on the FashionMNIST dataset for different replacement levels. The Y-axis is cut off between 0 and 2.5 for clearer presentation of the loss differences towards the end of the training. }
    \label{fig:eqx-fashionmnist-test-loss}
\end{figure}

\begin{figure}[ht!]
    \centering
    \includegraphics{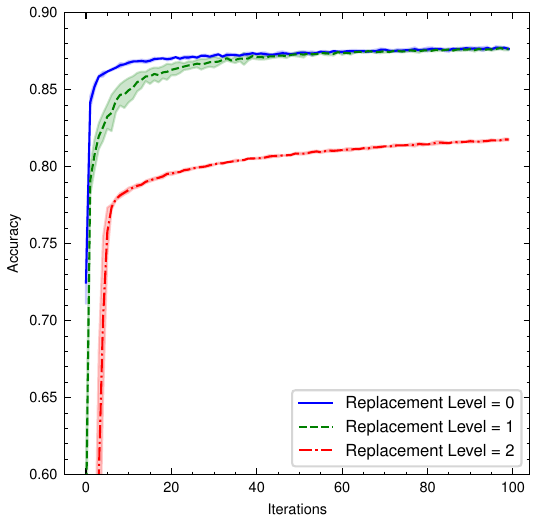}
    \caption{Accuracy plot during the training on the FashionMNIST dataset for different replacement levels. The Y-axis is cut off between 0.6 and 0.9 for clearer presentation of the accuracy differences towards the end of the training. }
    \label{fig:eqx-fashionmnist-train-acc}
\end{figure}

\begin{figure}[ht!]
    \centering
    \includegraphics{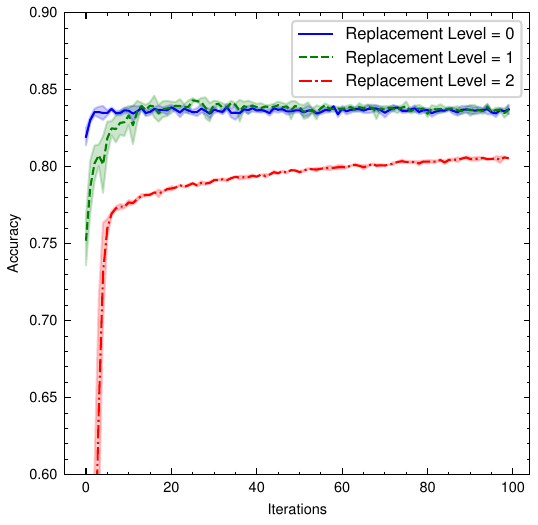}
    \caption{Accuracy plot during the testing stage on the FashionMNIST dataset for different replacement levels. The Y-axis is cut off between 0.6 and 0.9 for clearer presentation of the accuracy differences towards the end of the training. }
    \label{fig:eqx-fashionmnist-test-acc}
\end{figure}

\begin{figure}[ht!]
    \centering
    \includegraphics{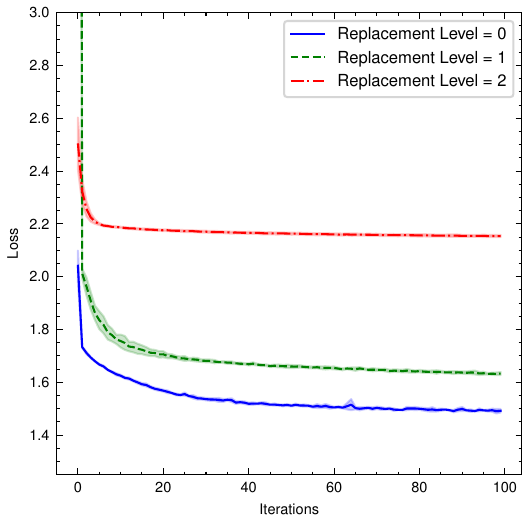}
    \caption{Loss plot during the training on the CIFAR-10 dataset for different replacement levels. The Y-axis is cut off between 1.25 and 3.0 for clearer presentation of the loss differences towards the end of the training. }
    \label{fig:eqx-cifar10-train-loss}
\end{figure}

\begin{figure}[ht!]
    \centering
    \includegraphics{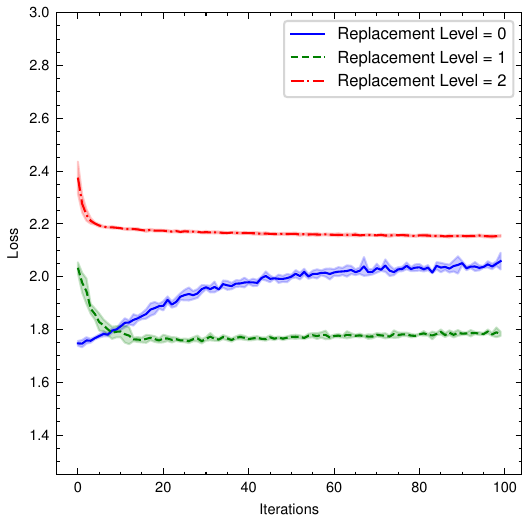}
    \caption{Loss plot during the testing stage on the FashionMNIST dataset for different replacement levels. The Y-axis is cut off between 1.25 and 3.0 for clearer presentation of the loss differences towards the end of the training. }
    \label{fig:eqx-cifar10-test-loss}
\end{figure}

\begin{figure}[ht!]
    \centering
    \includegraphics{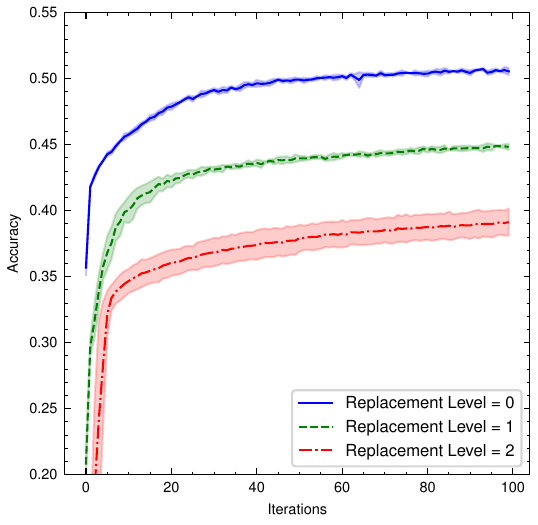}
    \caption{Accuracy plot during the training on the CIFAR-10 dataset for different replacement levels. The Y-axis is cut off between 0.2 and 0.55 for clearer presentation of the accuracy differences towards the end of the training. }
    \label{fig:eqx-cifar10-train-acc}
\end{figure}

\begin{figure}[ht!]
    \centering
    \includegraphics{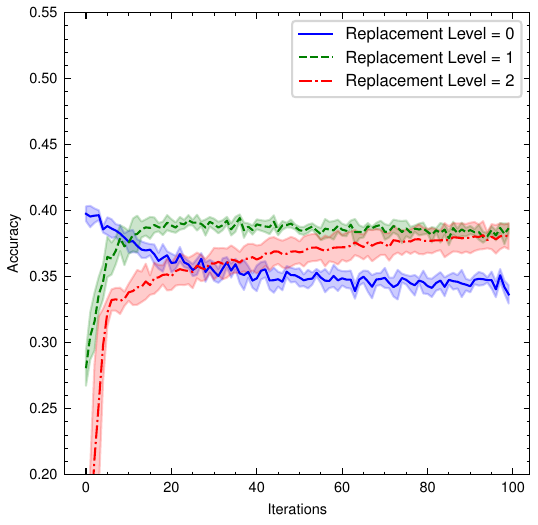}
    \caption{Accuracy plot during the testing stage on the CIFAR-10 dataset for different replacement levels. The Y-axis is cut off between 0.2 and 0.55 for clearer presentation of the accuracy differences towards the end of the training. }
    \label{fig:eqx-cifar10-test-acc}
\end{figure}

The average performance (loss and accuracy) plots of the JAX-based numerical experiments during the training and testing iterations are shown from Fig.~\ref{fig:eqx-mnist-train-loss} to Fig.~\ref{fig:eqx-cifar10-test-acc}. The performance curves are shaded, with the lower bound of the shade equals to \texttt{avg-std} and the upper bound of the shade equals to \texttt{avg+std}, where \texttt{avg} is the averaged performance metrics (loss and accuracy) and \texttt{std} is the standard deviation of the performance metrics, both over five repetitions.

\end{appendices}

\clearpage
\bibliography{sn-bibliography}

\end{document}